\documentclass[sigconf]{acmart}
\AtBeginDocument{%
  \providecommand\BibTeX{{%
    \normalfont B\kern-0.5em{\scshape i\kern-0.25em b}\kern-0.8em\TeX}}}

\usepackage{paracol}
\globalcounter{algocf}
\usepackage[bottom]{footmisc}
\usepackage{array}
\usepackage{wrapfig}
\usepackage{stmaryrd}

\newcolumntype{P}[1]{>{\centering\arraybackslash}p{#1}}
\usepackage{adjustbox}
\usepackage{graphicx}
\usepackage{algorithmic}
\usepackage{textcomp}
\usepackage{comment}
\usepackage{tikz}
\usetikzlibrary{automata, positioning, arrows, shapes.misc, patterns}
\usepackage{multirow}
\usepackage{enumitem}
\usepackage{graphics}
\usepackage[ruled,vlined]{algorithm2e}
\usepackage{url}
\usepackage{hyperref}
\usepackage{amsmath}
\usepackage{amsmath}
\usepackage{hyperref}
\usepackage[english]{babel}
\usepackage{caption}
\usepackage{subcaption}
\usepackage{multirow}
\usepackage{multicol}
\usepackage{cleveref}
\usepackage{graphicx}
\usepackage{mathtools}
\usepackage{lineno}
\usepackage{siunitx}
\usepackage{pifont}

\usepackage{ragged2e}
\newlength\myindent
\graphicspath{{figs/}}

\newcommand{\states}{\mathrm{\Sigma}}
\newcommand{\statespace}{\states}

\newcommand{\Globally}{\mathbf{G}}

\newcommand{\AP}{\mathit{AP}}
\newcommand{\trace}{t}

\newcommand{\zplus}{\mathbb{Z}_{\geq 0}}

\newcommand{\DD}{{d}}
\newcommand{\TT}{\mathbb{T}}

\newcommand{\HH}{\mathbf{H}}
\newcommand{\AAA}{\mathbf{A}}
\newcommand{\CC}{\mathbf{C}}
\newcommand{\ser}{\odot}

\newcommand{\Monitor}{$\mathtt{Monitor}$\xspace}
\newcommand{\Progress}{$\mathtt{Progress}$\xspace}
\newcommand{\Reduce}{$\mathtt{Reduce}$\xspace}
\newcommand{\THandler}{$\mathtt{Trace~ Handler}$\xspace}
\newcommand{\Aggregation}{$\mathtt{Aggregation}$\xspace}
\newcommand{\Counting}{$\mathtt{Counting}$\xspace}
\newcommand{\Operation}{$\mathtt{Aggregate~Operation }$\xspace}

\definecolor{bln_blue}{HTML}{00A8FF}
\definecolor{bln_red}{HTML}{c23616}
\definecolor{bln_green}{HTML}{16A085}
\definecolor{bln_magenta}{HTML}{9B59B6}
\definecolor{blond}{rgb}{0.98, 0.94, 0.75}
\definecolor{beige}{rgb}{0.96, 0.96, 0.86}
\definecolor{babyblueeyes}{rgb}{0.63, 0.79, 0.95}
	\definecolor{beaublue}{rgb}{0.74, 0.83, 0.9}

 \usepackage{algorithm2e}
\SetAlFnt{\normalsize}
\SetAlCapFnt{\normalsize}
\SetAlCapNameFnt{\normalsize}
\algsetup{linenosize=\normalsize}

\copyrightyear{2023} 
\acmYear{2023} 
\setcopyright{acmlicensed}\acmConference[MEMOCODE '23]{21st ACM-IEEE International Conference on Formal Methods and Models for System Design}{September 21--22, 2023}{Hamburg, Germany}
\acmBooktitle{21st ACM-IEEE International Conference on Formal Methods and Models for System Design (MEMOCODE '23), September 21--22, 2023, Hamburg, Germany}
\acmPrice{15.00}
\acmDOI{10.1145/3610579.3611075}
\acmISBN{979-8-4007-0318-8/23/09}

\begin{document}

\title{QTWTL: Quality Aware Time Window Temporal Logic for Performance Monitoring}

\author{Ernest Bonnah}

\affiliation{%
  \institution{University of Missouri-Columbia}
  \streetaddress{411 S 6th St}
  \city{Columbia}
  \state{Missouri}
  \country{USA}
  \postcode{65211}}
 \email{ernest.bonnah@mail.missouri.edu}

\author{Khaza Anuarul Hoque}
\affiliation{%
 \institution{University of Missouri-Columbia}
 \streetaddress{411 S 6th St}
 \city{Columbia}
 \state{Missouri}
 \country{USA}}
\email{hoquek@umsystem.edu}


\begin{abstract}
In various service-oriented applications such as distributed autonomous delivery, healthcare, tourism, transportation, and many others, where service agents need to perform serial and time-bounded tasks to achieve their goals, quality of service must constantly be assured. In addition to safety requirements, such agents also need to fulfill performance requirements in order to satisfy their quality of service. This paper proposes the novel quality-aware time window temporal logic (QTWTL) by extending the traditional time window temporal logic (TWTL) with two operators for counting and aggregation operations. We also propose offline runtime monitoring algorithms for the performance monitoring of QTWTL specifications. To analyze the feasibility and efficiency of our proposed approach, we generate a large number of traces using the New York City Taxi and Limousine Commission Trip Record data, formalize their performance requirements using QTWTL, and monitor them using the proposed algorithms. The obtained results show that the monitoring algorithm has a linear space and time complexity with respect to the number of traces monitored.
\end{abstract}





\maketitle

\section{Introduction}
\label{intro}
Monitoring of performance requirements has become increasingly necessary in distributed service-oriented systems such as delivery systems~\cite{shahzaad2022service}, tourism~\cite{mendes2016monitoring}, transportation~\cite{sodhro2019quality}, healthcare~\cite{guglielmetti2022improving}, and many other sectors. Agents in such systems are expected to perform numerous tasks within a fixed time-bound while complying with safety and performance requirements. (ISO 18646-3:2021, ISO 26262) 
For instance, for a group of autonomous delivery robots, a delivery task while adhering with the \emph{safety} requirement can be specified as ``\emph{deliver to location $A$ between time $T_1$ and $T_2$ and then deliver to address $B$ between time $T_3$ and $T_4$ while avoiding obstacles on the road"}. Consequently, a \emph{performance} requirement for the same set of robots can be specified as \emph{at least 90\% of the delivery robots should deliver the package within 30 minutes}. However, due to the dynamics of the real-world environment (traffic, weather, obstacles, and other uncertainties), there are cases where service-oriented agents may not be able to adhere to the performance requirements. Thus, there is a dire need to monitor the performance of these agents in addition to safety to ensure service quality. 

Temporal logic, specifically linear temporal logic (LTL), has been extensively used in formalizing temporal specifications in the control and analysis of robotic applications~\cite{cai2021modular, sahin2019multirobot, kloetzer2020path,xie2020temporal}. Even though LTL offers a rich set of operators, it also has a few limitations when it comes to robotic mission specifications with explicit time constraints. Such specifications can be expressed using bounded linear temporal logic (BLTL)~\cite{tkachev2013formula}, signal temporal logic (STL)~\cite{maler2004monitoring}, metric temporal logic (MTL)~\cite{saha2017task} and time window temporal logic (TWTL)~\cite{vasile2017time} which is a \textit{domain-specific} language for robotic systems specification. The TWTL language has rich semantics for specifying different time-bounded specifications and has several advantages. Specifically, time-bounded specifications can be represented more compactly and comprehensively in TWTL compared to BLTL, MTL, and STL. 
For instance, let us consider a specification as ``\textit{stay at P for 3 time steps within the time window [0, 5] and after this stay at Q for 2 time steps within the time window [6, 8]}”. This can be expressed in TWTL as \texttt{$[{\HH}^3 P]^{[0,5]} \ser [{\HH}^2 Q]^{[6,8]}$}. This same requirement can be expressed in MTL as ${\bigvee}_{i=0}^{5-3}({\Globally}_{[i,i+3]} P \land {\bigvee}_{j=i+3+6}^{i+3+8-2}{\Globally}_{[j,j+2]} Q)$.  As shown in this example, it requires only 5 operators for specifying this requirement using TWTL, whereas 11 operators are required to formalize the same requirement using MTL. This demonstrates the succinct characteristics of TWTL over MTL. With the increasing complexity of requirements, the complexity of MTL formulae will also grow, which makes the formal analysis of MTL formulae in complex robotic applications particularly expensive. Moreover, another advantage of TWTL over the STL, BLTL, and, MTL is the explicit \emph{concatenation operator} which is very useful in specifying the serial tasks that are very common in real-world service-oriented applications, such as in robotics~\cite{re2}, package delivery~\cite{re1}, vehicle routing~\cite{re3}, manufacturing \cite{mosca2019multi}, surveillance~\cite{bonnah2022runtime}, and so on.

Recently, several extensions of STL has been proposed for monitoring of performance requirements \cite{ma2020sastl, ma2017runtime, zhao2022astl}. However, motivated by the above-mentioned benefits of TWTL, in this paper, we propose an alternative temporal logic for formal specification of performance requirements. To be specific, we propose a novel quality aware time window temporal logic (QTWTL) by extending TWTL with two new operators for aggregation (can be augmented with maximum, minimum, and average operations) and counting. Indeed aggregation and counting are two important characteristics in specifying quality of service requirements in service-based applications. Using QTWTL we can specify interesting properties such as \emph{``the average time taken for robots to deliver a package should not exceed 5 time units"}. We also propose offline runtime monitoring algorithms for monitoring QTWTL specifications leveraging rewriting rules~\cite{rocsu2005rewriting} which evaluate formal specifications on finite execution traces online (and offline) by processing each event as it arrives. Runtime monitoring/verification (RV)~\cite{leucker2009brief} is a lightweight verification technique that checks whether a run of a system satisfies or violates a given correctness property, which is also a requirement for robotic or autonomous systems that are regulated by international standards (ISO 26262, IEC 61508) \cite{sanchez2019survey}. In contrast to online runtime monitoring~\cite{attard2021better,off4}, in offline runtime monitoring~\cite{off1,leucker2009brief,off3,off4}, the system is not directly monitored while executing. Instead, the relevant system events are recorded as an execution trace inside a data store. Once the monitored system terminates (or whenever a satisfactory number of events have been recorded), the collected execution trace is forwarded to the offline monitor. The offline monitor, which is entirely independent of the system, then proceeds by inspecting the system events recorded in the trace. Provided that the trace provides enough information, the monitor can deduce whether the correctness property was satisfied or violated. Offline monitoring is less intrusive than online monitoring, as it does not interfere with the system except for logging events, thus imposing little runtime overhead. This, however, comes at the cost of late detection since violations can only be detected once the system stops executing, which is when the monitor can inspect the recorded trace. Hence, offline monitoring is more suited for quality assurance/testing of large-scale deployed systems such as in our case. To demonstrate the feasibility and efficiency of our approach, we generate traces using the New York City Taxi and Limousine Commission Trip Record dataset~\cite{dataset}, formalize important performance requirements using QTWTL, and monitor them using the proposed algorithms. The obtained results show that the monitoring algorithm has a linear space and time complexity with respect to the number of traces monitored.

The rest of the paper is organized as follows: Section \ref{sec:prelim} presents the preliminaries. The syntax and semantics of QTWTL is then presented in Section \ref{qtwtl}. The proposed rewriting algorithms for QTWTL monitoring are introduced in Section \ref{offline_mon}. We evaluate the effectiveness of the proposed algorithms in Section \ref{case_study}. Related works is discussed in Section \ref{related_works}. Finally, Section \ref{conclusion} then concludes the paper.

\section{Preliminaries}
\label{sec:prelim}
Let $\AP$ be a finite set of {\em atomic propositions} and $\statespace= 2^{\AP}$ be the powerset of $\AP$. Let ($\zplus \times \statespace \times W$) be the {\em alphabet}, where $\zplus$ is the set of non-negative integers and $W$ is a set of real values for each parameter of interest (explained in the next section). A {\em time-stamped event} is a member of our alphabet and is of the form $(\tau, e, w)$, where $\tau \in \zplus$, $e \in \statespace$,  and $w \in W$. A {\em timed trace} (or simply {\em trace}) $\trace$ is a finite sequence of time-stamped events of the form $\trace =  (\tau_0, e_0, w_0), (\tau_1, e_1, w_1),  \dots (\tau_n, e_n, w_n)$, where $|\trace| = n+1$ and $n \in \zplus$. We abbreviate trace $\trace$ by $(\tau_i, e_i, w_i)_{i \in \zplus}$ and $i \leq n$. We call the sequence $\tau_0\tau_1\tau_2\cdots\tau_n$ {\em time-stamps} and their indices $i$ {\em time-points} and require that (1) $\tau_0 = 0$, and (2) $\tau_i < \tau_{i+1}$, for every $i \ge 0$; i.e., time-stamps are strictly monotonic. Let $\trace[i,j]$ denote the subtrace of $\trace$ from time-point $i \geq 0$ up to and including time-point $j \ge i$. For a trace $\trace$, we represent the ${i}^{th}$ time-stamped event by $\trace[i]$. Thus, for a trace $\trace = ( \tau_i, e_i, w_i)_{i \in \zplus}$, by $\trace[i].e$, we mean $e_i$ and by $\trace[i].\tau$, we mean $\tau_i$.

\begin{table*}[!t]
\centering
\caption{\label{Semantics}Semantics of QTWTL}
\normalsize
\begin{tabular}{|lll|}
\hline
$\TT \models \top$ & \text{iff} & $\forall t[i,j] \in \TT. ~ t[i,j] \models \top$ \\
$\TT \models \HH^{\DD} p $  & \text{iff} & $\forall t[i,j] \in \TT$.~\text{$p \in t[n].e$, $\forall n \in \{i,...,i + \DD\}$} \text{$ ~\wedge~ (t[j].\tau - $}  \text{$t[i].\tau) \geq \DD$} \\
$\TT \models \HH^{\DD} \neg p$  & \text{iff} & $\forall t[i,j] \in \TT$.~\text{$p \notin t[n].e$, $\forall n \in \{i,...,i + \DD\}$}  \text{$ \wedge~ (t[j].\tau - $}  \text{$ t[i].\tau) \geq \DD$} \\
$\TT \models  \phi_1 \wedge \phi_2$  & \text{iff} & $\forall t[i,j] \in \TT$.~\text{$(\trace[i, j] \models \phi_1) ~\wedge$}~ \text{$(\trace[i, j] \models \phi_2)$} \\
$\TT \models  \neg \phi $ & \text{iff} & $\forall t[i,j] \in \TT$.~\text{$\neg (\trace[i,j] \models \phi)$} \\
$\TT \models \phi_1 \ser\phi_2$ &  \text{iff} & $\forall t[i,j] \in \TT$. ~(\text{$\exists k = \arg\min_{i \le k \le j} \{\trace[i,k] \models \phi_1\}$}) $\wedge (\trace[k + 1,j] \models \phi_2)$ \\
$\TT \models [\phi]^{[a, b]}$   & \text{iff} & $\forall t[i,j] \in \TT$.~\text{$\exists k \geq i + a,~ s.t.~ $} \text{$\trace[k,i + b] \models \phi ~\wedge$} \text{$(t[j].\tau - t[i].\tau) \geq b$} \\
$\TT \models \CC (\phi) \sim c$ & \text{iff} &$\forall t[i,j] \in \TT$.~ \text{$\left( \frac{1}{|\TT|} \sum_{t[i,j] \in \TT}  \mathbb{1} (t[i,j] \models \phi]) \right) \sim c$}\\
$\TT \models \AAA_{\sigma} (h) \sim c$ & \text{iff} & $\forall t[i,j] \in \TT \cdot \alpha_{\sigma}^{h}(\{M_z \mid M_{z} \in M\}) \sim c$, \text{where $z \in [i,j]$} \\ \hline
\end{tabular}
\end{table*}

\section{Quality Aware Time Window Temporal Logic}
\label{qtwtl}
QTWTL extends TWTL~\cite{vasile2017time} with two new operators namely an aggregation operator ($\AAA$) and a counting operator ($\CC$). Based on the desired operation, the aggregation operator allows for the aggregation and evaluation of measurements of the same type (e.g. time, altitude, speed, etc.) generated across different devices. The counting operator enables the measuring of the percentage of traces generated from a set of devices that satisfy a given specification. 

Let a set of logged traces be denoted as $\TT=\{t_1,...,t_N\}$ where each $t_k \in \TT$ represents a trace and $1\le k \le N$. We assume that all traces in the trace set $\TT$ are synchronous, i.e., the time-stamps of all traces are aligned starting from time-point $i \geq 0$ up to and including time-point $j \geq i$. Recall that a trace $t_{k} \in \TT$ is of the form $( \tau_z, e_z, w_z)_{z \in \zplus}$ where $z \le N$. For each $t_{k} \in \TT$, there exists $W_k$ where each $w_{z} \in W_{k}$ is a real value for a parameter of interest, $h$, such as speed, distance covered, duration of trip, number of trips completed, rating, etc. at time-point $z$. We define $\pi_{z}^{h} (.)$ as a function that takes a set of logged traces and projects the value of $h$ at each time-point $z$ for each trace $\trace_k \in \TT$. Thus, for each $z \in [i,j]$, the set $M$ is created to contain the projected values of $h$ across $\TT$ such that $M = \{\{M_i\}, \{M_{i+1}\},..., \{M_j\}\}$. Each $M_{z} \in M$ then denotes a set of non-null real values of parameter $h$ at time-point $z$ across the trace set $\TT$. Given $M_{z} = \{m_{t_{1},z}, m_{t_2, z},..., m_{t_N, z}\}$, each $m_{{t_k}, z}, \in M_z$ is the projection of the real value  of $h$ in trace $\trace_k$ at time-point $z$ obtained through the function $\pi_{z}^{h} (.)$. The set of QTWTL formulas is inductively defined by the following grammar: 

\begin{equation*}
\begin{aligned}
&\varphi := \AAA_\sigma (h) \sim c \mid \CC (\phi) \sim c \mid \phi\\
&\phi := \top \mid \HH^{\DD} p \mid  \HH^{\DD} \neg p \mid \phi_1 \wedge \phi_2 \mid  \neg \phi \mid \phi_1 \ser \phi_2 \mid [\phi]^{[a,b]} \\
\end{aligned}
\end{equation*}

\vspace{2mm}
\noindent where $\top$ stands for true, $p$ is an atomic proposition in $AP$. The operators $\HH^{\DD}$, $\ser$ and $[\  ]^{[a,b]}$ represent the hold operator with $\DD \in \zplus$, concatenation operator and within operator respectively within a discrete-time constant interval $[a, b]$, where $a, b \in \zplus$ and $b \geq a$, respectively and $\wedge$ and $\neg$ are the conjunction and negation operators respectively. 
The $\AAA_{\sigma}$ and $\CC$ operators represent the \textit{aggregation} and \textit{counting} operators, respectively, $\sigma \in \{min, max, avg\}$ is the minimum, maximum and average respectively,  $\sim ~ \in \{\neq, <, >, \leq, \geq\}$, and $c \in \mathbb{R}$ is a constant. \\


The formula $\phi = \top$ always holds. With $\phi = \HH^{\DD} p$ and $\HH^{\DD} \neg p$, $p$ is expected to be repeated or not repeated for $\DD$ time units with the condition that $p \in t[n].e$ and $p \notin t[n].e$ respectively in all traces in $\TT$. With $\phi = \phi_1 \wedge \phi_2 $, all traces in $\TT$ must satisfy both formula while in  $\neg \phi$, no trace in $\TT$, satisfies  the given formula. A given formula in the form $\phi_1 \ser\phi_2$ specifies that every $\trace[i,j] \in \TT$ should satisfy the first formula first and the second afterward with one time unit difference between the end of execution of $\phi_1$ and start of execution of $\phi_2$. $\arg\min_{i \le k \le j}\{t[i,k] \models \phi_1\}$ simply returns the minimum time-point $k$ within the time bound $[i, k]$ for which $\phi_1$ is satisfied. The trace set $\TT$, must satisfy $\phi$ between the time window $[a, b]$ given $[\phi]^{[a, b]}$. The operator $\CC$ specifies the proportion of traces in $\TT$ that satisfy $\phi$ with respect to the total number of traces in $\TT$. The function $\mathbb{1}$  returns 1 when $(t[i,j] \models \phi])$ is true and 0 when otherwise. We denote $y \in \mathbb{R}$ as $y=\frac{|T_{sat}|}{|\TT|}$, where $|T_{sat}|$ is the summation of the satisfactions returned by $\mathbb{1}$ and $|\TT|$ is the total number of traces. The $\CC$ operator returns \textit{true} if and only if $y \sim c$ holds true. 
The $\AAA_{\sigma}$ operator performs an aggregation operation $\sigma$, e.g. \textit{max, min}, and \textit{average} over the set of traces reasoning on the parameter of interest. This is achieved using the function $\alpha_{\sigma}^{h} (.)$ which applies the aggregation operation on each ${M_{z}} \in M$. The $\AAA_{\sigma}$ operator returns \textit{true} if and only if for all $z \in [i,j]$, $\alpha_{\sigma}^{h}({M_{z}}) \sim c$ holds true. Formally, we define the semantics of QTWTL as the satisfiability relation $\TT \models \phi$ that relates timed-traces $\TT$ with the QTWTL formula $\phi$ in Table~\ref{Semantics}.\\

Below, we show some illustrative examples of QTWTL specifications and their natural language translations:
\begin{itemize}
   
\item \textit{At least $75\%$ of drivers should drop off passengers ($drop\_off$) at the drop off location within 35 time units of picking them up ($pick\_up$)}. This can be specified in QTWTL as $\varphi = \CC([\HH^1 \\ pick\_up] \ser [\HH^1 \ drop\_off]^{[0, 35]}) \geq 0.75$. \\



\item \textit{The average trip ratings ($trip\_rating$) for drivers should be more than 2 stars}. This can be specified as in QTWTL as $\varphi = \AAA_{avg} (trip\_rating) \geq 2$. 
\vspace{-2mm}
\end{itemize}

\section{Offline Monitoring Algorithms for QTWTL}
\label{offline_mon}
In this section, we discuss the overview and details of our proposed algorithms.
\subsection{Trace Preprocessing}
\label{trace_preprocessing}
In Section~\ref{sec:prelim}, we assume the traces are synchronous, and the sequence of time-stamps of those traces increases monotonically. In most cases, this assumption does not reflect real-world distributed systems. In real distributed systems, traces may be asynchronous with gaps in time-stamps, such as $t = (1, \{a, b\}, \{2\}), (3, \{b\},  \{4\}), \\ (5, \{a, b\}, \{3\})$. However, since we are performing offline monitoring, this allows us to utilize preprocessing of traces so that they suit our offline monitoring methods. Specifically, to fill in the gaps between the time-stamps in traces, we introduce silent events ($\epsilon$) for events and 0 for the parameter of interest. For instance, for the previous trace $t$, our preprocessing will generate a trace  $t' = (1, \{a, b\}, \{2\}), (2, \{\epsilon\}, \{0\}),(3, \{b\}, \{4\}), (4, \{\epsilon\}, \{0\})$, $(5, \{a, b\}, \{3\})$. Once all the traces are successfully preprocessed, we feed them to the monitoring algorithm, as presented in the following sections. 


\subsection{Algorithm Overview}
Our proposed QTWTL monitoring algorithm consists of the following five parts:
\begin{itemize}
    \item \Monitor: The monitor outlines the overall runtime monitoring approach of a QTWTL formula. Given a set of traces $\TT$ and a QTWTL formula $\phi$, the monitor checks if $\TT$ $\models$ $\phi$.
    \item \THandler: For each $t \in \TT$, the \THandler iterates over the new events by recursively rewriting (inspired by the rewriting-based approach for LTL introduced in~\cite{rocsu2005rewriting}) the given formula using the \Progress and \Reduce algorithms. The \THandler terminates if there are no new events or the re-written formula evaluates to \textit{true} or \textit{false} indicating the satisfaction or violation of the given formula, respectively.   
    \item \Progress: For each event $e_i \in t$, the progress algorithm rewrites the given formula $\phi$  to a new formula $\psi$ based on the QTWTL semantics (Section II). 
    \item \Reduce: The rewritten formula $\phi$ is reduced to an equivalent formula $\beta$ using the propositions of various QTWTL formulas.
    \item \Counting: Given a QTWTL requirement $\phi$ and set of traces $\TT$, the \Counting algorithm counts the number of traces that satisfy the requirement.
    \item \Aggregation: The \Aggregation algorithm allows for monitoring of parameters of interest expressed as QTWTL specification $\phi$ across different traces using \Operation. 
    \item \Operation: The \Operation algorithm details the computation of aggregation operations given $\sigma \in \{min, max, avg\}$. 
\end{itemize} 

\subsection{Algorithm Details}
In this section, we discuss the five parts of our proposed algorithm in detail. \\

\setlength{\textfloatsep}{3pt}
\SetInd{0.5em}{0.5em}
\begin{algorithm}[!t]
	\caption{\Monitor}
	\label{alg:dse}
        \DontPrintSemicolon
    
        \textbf{Function:} \Monitor $(\phi, \TT)$ \\
        \SetKwInOut{Input}{Inputs}\SetKwInOut{Output}{Outputs}
        \Input{QTWTL formula $\phi$, Trace Set $\TT$}
		\Output{Verdict = {$\bot, \top$}}
        \begin{algorithmic}[1]
        \IF{$\phi = \AAA_{\sigma} (h) \sim c$}       
        \STATE $\beta \leftarrow$ \Aggregation$(h, \sim, c, \sigma, \TT)$
        
        \ENDIF
        \IF{$\phi = \CC (\phi) \sim c$} 
            \STATE $\beta \leftarrow$  \Counting$(\phi, \TT, \sim, c)$
        \ENDIF
        \IF{$\phi \in \{ \HH^{\DD} p,  \HH^{\DD} \neg p,  \phi_1 \wedge \phi_2, \neg \phi, \phi_1 \ser \phi_2,$ \\ ~~~~~~~~~~$[\phi]^{[a,b]} \}$}
        \STATE $\beta \leftarrow \top$
            \FOR{$\trace \in \TT$}
            \IF {(\THandler$(\phi, \trace)$ == $\bot$)}
                \STATE $\beta \leftarrow \bot$
                \STATE \textbf{break}
                \ENDIF
            \ENDFOR
        \ENDIF
        \IF{$\phi = \_$} 
            \RETURN $\bot$
        \ENDIF
        \IF{$\TT = \{~\}$} 
            \RETURN $\bot$
        \ENDIF
        \RETURN $\beta$
        \end{algorithmic}
\end{algorithm}

\SetInd{0.5em}{0.5em}
\begin{algorithm}[!t]
	\caption{\THandler}
	\label{alg:dse}
        \DontPrintSemicolon
        
        \textbf{Function:} \THandler $(\phi, \trace)$ \\
        \SetKwInOut{Input}{Inputs}\SetKwInOut{Output}{Outputs}
        \Input{QTWTL formula $\phi$, Trace $\trace$}
		\Output{Verdict = {$\bot, \top$}}

        \begin{algorithmic}[1]
        \STATE $\beta \leftarrow \mathtt{Reduce}(\phi)$
         \IF{$\beta \in \textcolor{black}{\{\top, \bot\}}$}
           \STATE \textbf{break}
         \ELSE
            
            \WHILE{get\_event $(e_i \in t)$}
              \STATE $\psi \leftarrow \mathtt{Progress}(\phi, e_i)$
              \STATE $\beta \leftarrow \mathtt{Reduce}(\psi)$ 
                \IF{$\beta \in \textcolor{black}{\{\top, \bot\}}$}
                  \STATE \textbf{break}
                \ELSE
                   \STATE $\phi \leftarrow \beta$
                \ENDIF
             \ENDWHILE
            
         \ENDIF     
        \RETURN $\beta$
		\end{algorithmic}
\end{algorithm}

\indent \textbf{Monitoring Algorithm:}
Given a set of traces $\TT$ and a QTWTL formula $\phi$, the \Monitor algorithm (Algorithm 1) returns a Boolean value indicating whether the set of traces satisfies the given QTWTL specification or not. The \Monitor does this by first checking the structure of the specification and subsequently calling the appropriate algorithm to evaluate the satisfaction or violation of the specification. The algorithm calls \Aggregation algorithm if the given formula has the $\AAA_{\sigma}$ operator (Lines 1-3). The \Counting algorithm is called for the evaluation of the formula if the formula has the form $\CC(\phi) \sim c$  (Lines 4-6). If the given QTWTL formula has any other form involving $\HH, \land, \lor, \ser, \neg, [~]^{[a, b]}$, the \THandler algorithm is recursively called on each $\trace \in \TT$ to evaluate the satisfaction of the specification. The iteration over $\TT$ is stopped if any $\trace \in \TT$ violates the specification (Lines 7-15). The default condition is represented by the case in (Lines 16-17) for which we return $\bot$. The algorithm returns $\bot$ if an empty $\TT$ is passed as an input to it (Lines 19-21).\\

\indent \textbf{Trace Handler Algorithm:}
The \THandler algorithm (Algorithm 2) iterates through time-stamped events in trace $\trace$ until a verdict is obtained. The algorithm first attempts to reduce the input formula to check for a satisfaction or violation (Lines 1-3). If the formula does not yield a satisfaction or violation after the initial reduction, a time-stamped event is read from the trace (Line 5) after which the \Progress function is invoked to rewrite $\phi$ according to the event $e_i$ and QTWTL semantics (Line 6). The \Reduce function is then called to simplify the rewritten formula $\psi$ and check for the satisfaction or violation (Line 7).  The monitoring is terminated and a verdict returned either as satisfaction/violation ($\top / \bot$) or the monitoring continues for the incoming events. \\

\SetInd{0.5em}{0.5em}
\begin{algorithm}[!t]
	\caption{\Progress}
	\label{alg:dse}
		\normalsize
        \DontPrintSemicolon


        \begin{algorithmic}[1]
        \STATE $ $\Progress$(\phi,e_i) := $ 
        \STATE \textbf{match} \ $\phi$ \textbf{with} 
            \STATE  $\mid  \top \rightarrow  \top$ 
            \STATE  $\mid  \bot \rightarrow  \bot$ 
            \STATE  $\mid \HH^{\DD} p \rightarrow$  \textbf{match} $\DD$ \textbf{with} 
            \STATE  \ \ \ \ \ \ \ \ \ \ \ \ \ \ \ $\mid 0  \rightarrow$ \textbf{if} $p \ \in e_i$ \textbf{then} $\top$ \textbf{else}  $\bot$    
            \STATE \ \ \ \ \ \ \ \ \ \ \ \ \ \ \ $\mid 1 \rightarrow$ \textbf{if} $p \ \in e_i$ \textbf{then} $\top$ \textbf{else}  $\bot$
            \STATE \ \ \ \ \ \ \ \ \ \ \ \ \ \ \ $\mid \_ \rightarrow$ \textbf{if} $p \ \notin e_i$ \textbf{then} $\bot$ \textbf{else}  $ \ \HH^{\DD - 1} p$
            
            \STATE  $\mid \HH^{\DD} \neg p \rightarrow$  \textbf{match} $\DD$ \textbf{with} 
            \STATE  \ \ \ \ \ \ \ \ \ \ \ \ \ \ \ $\mid 0  \rightarrow$ \textbf{if} $p \ \notin e_i$ \textbf{then} $\top$ \textbf{else}  $\bot$  
            \STATE \ \ \ \ \ \ \ \ \ \ \ \ \ \ \ $\mid 1 \rightarrow$ \textbf{if} $p \ \notin e_i$ \textbf{then} $\top$ \textbf{else}  $\bot$
            \STATE \ \ \ \ \ \ \ \ \ \ \ \ \ \ \ $\mid \_ \rightarrow$ \textbf{if} $p \ \in e_i$ \textbf{then} $\bot$ \textbf{else}  $ \HH^{\DD - 1} \neg p$
            \STATE $\mid \ \phi_1 \, \land \, \phi_2 \rightarrow   $\Progress$ (\phi_1, e_i) \, \land \,  $\Progress$ (\phi_2, e_i)$
            \STATE $\mid \ \neg \phi \rightarrow  \ \neg \ $\Progress$(\phi, e_i)$ 
           
            \STATE $\mid \ \phi_1 \ser \phi_2 \rightarrow$  \textbf{match} \ \Progress$(\phi_1, e_i)$ \textbf{with} 
            \STATE \ \ \ \ \ \ \ \ \ \ \ \ \ \ \ \ \ \ \ \ \ $\mid \top \rightarrow 
            \ $\Progress$(\phi_2, e_i)$
            \STATE \ \ \ \ \ \ \ \ \ \ \ \ \ \ \ \ \ \ \ \ \ $\mid \_ \rightarrow ($\Progress$(\phi_1, e_i) \ser \phi_2)$
            
            \STATE $\mid \ [\phi]^{[a, b]} \rightarrow$ \textbf{match} \ $\phi$ \textbf{with}  
            \STATE \ \ \ \ \ \ \ \ \ \ \ \ \ \ \ \ \  $\mid \HH^{\DD} \, p \rightarrow $ \textbf{if} $\DD \leq (b- a)$ \textbf{then} 
            \STATE \ \ \ \ \ \ \ \ \ \ \ \ \ \ \ \ \ \ \ \ \ \ \ \ \ \ \ \ \  $ ($\Progress$ (\phi, e_i) \, \lor \, [\phi]^{[a+1, b]}) \ \textbf{else} \  \bot$ \vspace{5pt}
            \STATE \ \ \ \ \ \ \ \ \ \ \ \ \ \ \ \ \  $\mid \HH^{\DD} \, \neg p \rightarrow $ \textbf{if} $\DD \leq (b- a)$ \textbf{then} 
            \STATE \ \ \ \ \ \ \ \ \ \ \ \ \ \ \ \ \ \ \ \ \ \ \ \ \ \ \ \ \  $ ($\Progress$ (\phi, e_i) \, \lor \, [\phi]^{[a+1, b]}) \ \textbf{else} \  \bot$ \vspace{5pt} 
            \STATE \ \ \ \ \ \ \ \ \ \ \ \ \ \ \ \ \ $\mid \_ \rightarrow \ ($\Progress$(\phi, e_i) \lor [\phi]^{[a + 1, b]})$
            
        \end{algorithmic}
\end{algorithm}

\indent \textbf{Progress Algorithm:}
The \Progress \ is a recursive function\footnote{Note that Algorithm 3 and Algorithm 4 are described in the functional programming/Ocaml style.} (Algorithm 3) that takes QTWTL formula and an event as input and returns the corresponding rewritten formula based on the QTWTL semantics (Section II). The first two cases (Line 1 and Line 2) are trivial (i.e., false and true) and \Progress function returns false and true, respectively. In the rewriting of the hold operator $\HH^{\DD} p$  (Line 5), $\HH^{0} p = \HH^{1} p = p$. The \Progress Algorithm returns true or false after checking whether the proposition $p$ is in the event $e_i$ or not for the cases in Lines 6 and 7. For the case where $\DD > 1$, if $p$ is in the current event $e_i$, the algorithm returns $\HH^{\DD - 1} p$ to show the outstanding holding condition $\DD$ required for $p$ to be satisfied in the upcoming events (Line 8). The rewriting for the hold operator with the negation of a given proposition (i.e., $\HH^{\DD} \neg p$) is evaluated as the previous case (Lines 9-12). In the case of the conjunction operator (Line 13), the \Progress function is recursively applied on the operands (i.e., $\phi_1$ and $\phi_2$). For the negation operator (Line 14), we push the \Progress function inside the negation operator. The rewriting of the concatenation operator ($\phi_1 \ser \phi_2$) is performed based on two cases (Lines 15-17): 1) if $\phi_1$ is true, the \Progress function is recursively applied on the formula $\phi_2$ and returns \Progress ($\phi_2$); 2) for all other cases, the \Progress function is recursively applied on the formula $\phi_1$ and returns \Progress$(\phi_1, e_i) \, \ser \, \phi_2$. \\ 
\indent Similarly, two cases are considered in the evaluation of the within operator (Line 18): 1) if the formula $\phi$ corresponds to the hold operator (Lines 19-22) then it is required that the holding condition $\DD$ is satisfied within the interval $[a, b]$. A disjunctive formula demonstrating the formula $\phi$ is satisfied in the current event $e_i$ or  $\phi$ will be satisfied within the remaining time interval $[a + 1, b]$ is returned. 2) for all other cases (Line 23), a disjunctive formula is returned as the previous case without any holding condition. \\

\indent \textbf{Reduce Algorithm:} The \Reduce is a recursive function (Algorithm 4) that takes a QTWTL formula as input and returns a reduced QTWTL formula. The \Reduce \ algorithm is composed of the following parts. The first two cases (i.e., $\neg$, $\land$) reduce the formula based on the Boolean simplification rules (Lines 3-12). The reduction of the concatenation operator ($\phi_1 \ser \phi_2$) requires following sub-cases: a) if one of the formulae is false ($\bot$) the algorithm returns false (Lines 13-17); b) if $\phi_1$ is true ($\top$) then the reduced form of the $\phi_2$ is returned, i.e., $ $\Reduce$(\phi_2)$ (Line 16); and c) in all other cases, we rerun $ $\Reduce$(\phi_1) \, \ser \, $\Reduce$(\phi_2)$ (Line 17). 
In the latter case for the within operator, $[\phi]^{[a, b]}$ is reduced to $\phi$ if $a = b$, i.e., unity interval. For all other case we return the formula without any reduction (Line 18). Note, in all the cases, $f$ and $g$ are denotations of a QTWTL formula.
Finally, If $\psi$ does not match any of the patterns between Lines 3-18, then Line 19 returns the formula.\\

\begin{algorithm}[!t]
	\caption{\Reduce}
	\label{alg:dse}
        \DontPrintSemicolon
        \begin{algorithmic}[1]
         \STATE $ $\Reduce$(\psi) := $ 
        \STATE \textbf{match} \ $\psi$ \textbf{with} 
            \STATE $\mid \neg f \rightarrow $ \textbf{match} $($\Reduce$ \ f)$ \textbf{with}
            \STATE \ \ \ \ \ \ \ \ \ \ \ \ \ \ $\mid \bot \ \ \rightarrow \top$ 
            \STATE \ \ \ \ \ \ \ \ \ \ \ \ \ \ $\mid \top \ \ \rightarrow \bot$ 
            \STATE \ \ \ \ \ \ \ \ \ \ \ \ \ \ $\mid \neg g \  \rightarrow \ g$ 
            
            \STATE $\mid \phi_1 \land \phi_2  \rightarrow$ \textbf{match} $($\Reduce$ \ \phi_1), ($\Reduce$ \ \phi_2)$ \textbf{with}  
            \STATE \ \ \ \ \ \ \ \ \ \ \ \ \ \ \ \ \ \ \ \ $\mid (\bot, f) \rightarrow \bot$      
            \STATE \ \ \ \ \ \ \ \ \ \ \ \ \ \ \ \ \ \ \ \ $\mid  (f, \bot) \rightarrow \bot$ 
            \STATE \ \ \ \ \ \ \ \ \ \ \ \ \ \ \ \ \ \ \ \  $\mid  (\top,  f) \rightarrow f$ 
            \STATE \ \ \ \ \ \ \ \ \  \ \ \ \ \ \ \ \ \ \ \ $\mid (f,  \top) \rightarrow f$ 
            \STATE \ \ \ \ \ \ \ \ \  \ \ \ \ \ \ \ \ \ \ \ $\mid (\top, \top) \rightarrow \top$ 
                       \STATE $\mid \phi_1 \ser \phi_2  \rightarrow$ \textbf{match} $($\Reduce$ \ \phi_1), ($\Reduce$ \ \phi_2)$ \textbf{with}
            \STATE \ \ \ \ \ \ \ \ \ \ \ \ \ \ \ \ \ \ \ \ $\mid \bot,  f \rightarrow \ \bot$ 
            \STATE \ \ \ \ \ \ \ \ \ \ \ \ \ \ \ \ \ \ \ \ $\mid f,  \bot \rightarrow \ \bot $
            \STATE \ \ \ \ \ \ \ \ \ \ \ \ \ \ \ \ \ \ \ \ $\mid \top, f \rightarrow \ f $
             \STATE \ \ \ \ \ \ \ \ \ \ \ \ \ \ \ \ \ \ \ \ $\mid f, g \rightarrow \ f \ser g $
        
            \STATE $\mid [\phi]^{[a, b]} \rightarrow$ \ \textbf{if} $a=b$ \textbf{then}  $\phi$ \textbf{else} $[\phi]^{[a, b]}$
            \STATE $\mid \_ \rightarrow \psi$
        \end{algorithmic}
\end{algorithm} 

\begin{algorithm}[t]
	\caption{\Counting}
	\label{alg:dse}
        \DontPrintSemicolon
        \textbf{Function:} \Counting$(\phi, \TT, \sim, c)$ \\
        \SetKwInOut{Input}{Inputs}\SetKwInOut{Output}{Outputs}
        \Input{$\phi$, $\TT$, $\sim$, $c$}
		\Output{Verdict = {$\bot, \top$}}
        \textbf{Initialization:} ${v := 0}$, \ ${n := {|\TT|}}$ \\
        
        \begin{algorithmic}[1]
                \FOR{$t \in \TT$}
                \IF {(\THandler$(\phi, \trace)$ == $\top$)}
                \STATE $v := v + 1$
                \ELSE  
                \STATE $v := v$ 
                \ENDIF
                \ENDFOR
                \STATE $v := v/n $ 
                \STATE  $\beta \leftarrow v \sim c$
            \RETURN $\beta$
        \end{algorithmic}
\end{algorithm}

\indent \textbf{Counting Algorithm:}
The \Counting Algorithm (Algorithm 5) counts the number of satisfaction over the set $\TT$ given a QTWTL formula. The algorithm then returns a verdict based on a comparison between the number of satisfactions and a constant $c$. For every $\trace \in \TT$, Algorithm 2 is called to verify the satisfaction or violation of the formula $\varphi$ (Lines 1-2). Consequently, for every $\trace \models \phi$, $v$ is incremented by 1 otherwise the value of $v$ is maintained (Lines 3-6). In Line 8, the average of the number of satisfaction is computed as the proportion of the number of satisfactions to the total number of traces. The computed value $v$ is subsequently compared to the constant $c$ (Line 9-10). \\ 

\indent \textbf{Aggregation Algorithm:} The \Aggregation Algorithm (Algorithm 6) takes as input a set of traces $\TT$, the parameter of interest $h$, operator $\sim$, the constant $c$, the aggregation operation $\sigma$, and returns a verdict based on the comparison between the aggregated value $x_z$ and a constant $c$. In Line 1, the algorithm based on the set of traces $\TT$ populates the set of sets $M$. This population is done using the function $\pi_{z}^{h}(.)$ based on the assumption that all traces in $\TT$ are bounded within $[i,j]$. Thus, for each $z \in [i,j]$, the set $M_z \in M$ is created to contain the values of the parameters of interest across $\TT$ such that $M = \{\{M_i\}, \{M_{i+1}\},..., \{M_j\}\}$. 
For each $M_z \in M$, the desired $AggregationOperation(.)$ is performed on the set $M_{z}$ as presented in Algorithm 7 using the function $\alpha_{\sigma}^{h}$. The resulting value $x_z$ is then compared to the constant $c$ (Lines 2-4). The algorithm breaks and returns $false$ if any $x_z \sim c$ does not hold true otherwise the algorithm continues and returns $true$ if for all $z \in [i,j]$, $x_z \sim c$ holds true (Lines 5-9). \\

\begin{algorithm}[t]
	\caption{\Aggregation}
	\label{alg:dse}
        \DontPrintSemicolon
        \textbf{Function:} \Aggregation$(h, \sim, c, \sigma, \TT)$ \\
        \SetKwInOut{Input}{Inputs}\SetKwInOut{Output}{Outputs}
        \Input{$h, \sim, c, \sigma, \TT$}
		\Output{Verdict = {$\bot, \top$}}
       \textbf{Initialization:} $M = \{\}$, $z:=0$ \\
        \begin{algorithmic}[1]
        \STATE $M \leftarrow \texttt{populate}(\TT)$
       
        \FOR{\textbf{each} $M_z \in M$}
        \STATE $x_z \leftarrow \alpha_{\sigma}^{h}(M_{z})$
        \STATE $\beta \leftarrow x_z \sim c$  
        \IF{$\beta == \bot$}
              \STATE \textbf{break}
        \ENDIF
        \ENDFOR
        \RETURN $\beta $
        \end{algorithmic}
\end{algorithm}

\begin{algorithm}[!t]
	\caption{\Operation}
	\label{alg:dse}
        \DontPrintSemicolon
        
        \textbf{Function:}~\textbf{$\alpha_{\sigma}^{h}(M_{z})$} \\
        \SetKwInOut{Input}{Inputs}\SetKwInOut{Output}{Outputs}
        \Input{$M_{z}, \sigma$}
		\Output{Aggregated Value $x_z$}
        \textbf{Initialization:} ${x_z := 0}$ \\
        \begin{algorithmic}[1] 
        \IF{$\sigma = min$}
            \STATE $x_z \leftarrow min(M_z)$
        \ENDIF
        \ \IF{$\sigma = max$}
            \STATE $x_z \leftarrow max(M_z)$
        \ENDIF
         \IF{$\sigma = avg$}
            \STATE $x_z \leftarrow avg(M_z)$
        \ENDIF
            \RETURN $x_z$
        \end{algorithmic}
\end{algorithm}

\indent \textbf{Aggregate Operation Algorithm:}
The \Operation Algorithm (Algorithm 7) performs the desired aggregation operation on a set  $M_{z}$. The algorithm based on the operation $\sigma \in \{min, max, avg\}$ computes the value $x_z$ which is subsequently compared to the given constant $c$ in Algorithm 6. In the first case, if $\sigma = min$, the minimum among all $m_{\trace_k, z} \in M_{z}$ is computed and returned (Lines 1-3). Similarly, if $\sigma = max$, the maximum among all $m_{\trace_k, z} \in M_{z}$ is computed and returned (Lines 4-6). In computing the average if $\sigma = avg$, the average of all the elements in $M_{z}$ is computed and subsequently returned as $x_z$ (Lines 7-9). \\ 

\indent \textit{Correctness}. We prove the correctness of the proposed monitoring algorithm using the following theorem. We prove this theorem based on the structural induction on the formula. 
Let $\phi$ be a QTWTL formula, then for any set of traces $\TT$, Algorithm 1 returns $\bot / \top$ \emph{iff} 
$[ \TT \models \phi ] = \bot / \top$. \\

\indent \emph{Proof sketch}: The base cases are associated with $\phi \in$ ($\top, \bot, \HH^{\DD} p, \HH^{\DD} \neg p $). The proof for base cases ($\top, \bot$) is trivial since Algorithm 3 mimics the semantics of QTWTL for $\top$ and $\bot$. The QTWTL semantics of the hold operators follows the case analysis presented in Algorithm 3. For Algorithm 1 to be correct, then 
$\TT \models \HH^{\DD} p \equiv$ \Monitor($\HH^{\DD} p, \ \TT)$. We perform a case analysis on $\phi$ while assuming all traces in set $\TT$ have a fixed size. Given $\trace[i,j] \in \TT$ and \textbf{$|\trace[i,j]| = 1$}, we now perform further case analysis on the value of $\DD$. 
Let us assume $\DD = 0$ or $\DD = 1$. Based on QTWTL semantics $\forall \trace[i,j] \in \TT \cdot \trace[i,j] \models \HH^{\DD} p$  \text{iff} \text{$p \in e_i, \forall n \in \{i\} \ \wedge$} \text{$(t[j].\tau - t[i].\tau) \geq \DD$}. With \text{$(t[j].\tau - t[i].\tau) \geq \DD$}, the above specification will be satisfied if $p \in e_i$ as stated above. Hence, $[\trace[i, j] \models \HH^{\DD} p]$ =$\top$ if $p \in e_i$ else $\bot$. According to Algorithm 1
\Monitor ($\HH^{\DD}p, \TT) = \top$ if all traces in $\TT$ satisfies the formula  $\HH^{\DD}p$. For each $\trace[i,j] \in \TT$, we call Algorithm 2 to check if $\trace[i,j]$ satisfies $\phi$ or not. After applying Algorithm 3, i.e., \Progress($\HH^{\DD}p, e_i)= \top$ if $p \in e_i$ else $\bot$. The same approach can be adopted in analyzing the case where $\DD > 1$. The formula $\HH^{\DD} \neg p$ is handled as the case above.

We proceed with the induction on the structure of the formula $\varphi \in \{\phi_1 \lor \phi_2$, $~\phi_1 \land \phi_2,~\phi_1 \ser \phi_2,~\neg \phi,~[\phi]^{[a, b]}\}$. The induction hypothesis requires that given a QTWTL formula and a set of traces, the theorem based on the recursion in Algorithm 1 returns $\bot / \top$ demonstrating whether the QTWTL formula is satisfied or not. In the case $\varphi = \phi_1 \land \phi_2$, for Algorithm 1 to be correct, then $\TT \models \phi_1 \land \phi_2 \equiv$ \Monitor($\phi_1 \land \phi_2, \TT)$. Based on QTWTL semantics, $\TT \models \phi_1 \land \phi_2$ = ($\TT \models \phi_1) \land (\TT \models \phi_2$). Likewise, from Algorithm 1, \Monitor($\phi_1 \land \phi_2, \TT)$ = \Monitor($\phi_1, \TT)) ~ \land$ \Monitor($\phi_2, \TT)$). Performing the induction on  $\phi_1$, i.e. $\phi_1 = \top$, $\varphi = \top \land \phi_2$. From the QTWTL semantics on the $\land$ operator, $\forall \trace[i,j] \in \TT \cdot \trace[i, j] \models (\top \land \phi_2) = \phi_2$. Likewise, from Algorithm 1, \Monitor($\top \land \phi_2, \TT) = \phi_2$, by applying Algorithm 2 on each $\trace[i,j] \in \TT$, \Reduce($\top \land \phi_2$) = $\phi_2$. The induction on $\phi_2$ can be performed in a similar manner to prove the correctness of Algorithm 1. Similarly, the induction on the concatenation and within operators are consistent with the theorem because all the cases analyzed in the various scenarios in Algorithms 2, 3, and 4 satisfy the semantics of QTWTL language.

In the case of the Counting operator, $\CC(\phi) \sim c$, Algorithm 1 is correct if [$\TT \models \CC(\phi) \sim c] \equiv$ \Monitor($\CC(\phi) \sim c, \ \TT)$. Based on the QTWTL semantics, the function $\rho(.)$ counts the number of traces from the set $\TT$ that satisfy the formula $\phi$ after evaluating each $\trace[i,j] \in \TT$ against the given QTWTL formula. From the semantics [$\TT \models \CC(\phi) \sim c] = \top$ if $y \sim c = \top$ else $\bot$, where $y = \mathbb{R}$ is the output of the function $\rho(.)$. According to Algorithm 1, \Monitor ($\CC(\phi) \sim c, \TT) = \top$ after applying Algorithm 5, i.e. \Counting$(\phi, \TT, \sim, c) = \top$ else $\bot$. For the Aggregation operator, $\AAA_\sigma (h) \sim c$, Algorithm 1 is correct if [$\TT \models \AAA_\sigma (h) \sim c] \equiv$ \Monitor($\AAA_\sigma (h) \sim c, \ \TT)$. Based on the QTWTL semantics, [$\TT \models \AAA_\sigma (h) \sim c] = \top$ if $\forall z \in [i,j]$, $\alpha_{\sigma}^{h}(M_{z}) = \top$ else $\bot$. According to Algorithm 1, \Monitor($\AAA_\sigma (h) \sim c, \ \TT) = \top$ after applying Algorithm 6, i.e. \Aggregation$(h, \sim, c, \sigma, \TT) = \top$ else $\bot$.  Note, \Aggregation$(h, \sim, c, \sigma, \TT) = \top$ if forall $z \in [i,j]$, the comparison between the aggregated parameter of interest generated by Algorithm 7 $x_z$ and the constant $c$ holds true, i.e., $x_{z} \sim c = \top$, where $x_{z} = \alpha_{\sigma}^{h}(M_{z})$.\\

\indent \textit{Complexity}. Given each trace in $\TT$ has a fixed length of $|t|$, the time complexity of Algorithm 1 is $\mathcal{O}(|\TT|)$.  Similarly, the space required for executing Algorithm 1 depends on the total number of traces in $\TT$. Thus, we can deduce that the space complexity is also $\mathcal{O}(|\TT|)$ for a trace set $\TT$ each with a fixed length of $|t|$. Given a fixed length of trace set, with the increasing number of events in each trace, the time complexity of Algorithm 1 is $\mathcal{O}(|\TT|+(|t|_{max}))$ where $|t|_{max}$ is the maximum length of events defined by the temporal operators of $\phi$. When there are two or more operators nested in a given formula, the time complexity for each operator is bounded by $\mathcal{O}((\phi)+ |t|_{max}))$. 
\section{Case Study}
\label{case_study}
In this section, we evaluate the feasibility of the proposed runtime monitoring algorithms for QTWTL specifications using a set of finite traces generated using the 2021 Yellow Taxi Trip Records dataset~\cite{dataset}. These records contain 2.82 million rows and 18 columns and are generated from the trip record submissions made by yellow taxi Technology Service Providers (TSPs). Each row represents a single trip in a yellow taxi in 2021. The trip records include fields capturing pick-up dates/times ($tpep\_pickup\_datetime$), drop-off dates/times ($tpep\_dropoff\_datetime$), pick-up locations ($PULocationID$), drop-off locations ($DOLocation ID$), trip distances ($trip\_distance$), itemized fares (tax ($mta\_tax$), toll ($tolls\_amt$), congestion surcharge ($cong\_surcharge$), improvement surcharge ($impr$\\$ovement\_ surcharge$), tip ($tip\_amount$), and total amount ($total\_am$\\$ount$)), rate types ($RatecodeID$), payment types ($payment\_type$), and driver-reported passenger counts ($passenger\_count$). 
We augmented the dataset with five parameters of interest, specifically, $taxi\_request\_time$ ($arrival\_time - m$ where $m$ is randomly varied in the range [1, 10]), $arrival\_time$ (pickup time$ - n$, where $n$ is randomly varied in the range [1, 3]), $wait\_time$ (difference between  $taxi\_request\_time$ and $arrival\_time$),  $pickup\_delay$ (difference between  $tpep\_pickup\_datetime$  and $arrival\_time$), and $trip\_rating$ (randomly generated in the range 1-5 stars) as this information is not available in the original dataset due to privacy reasons. We developed a Python program that reads the 2021 Yellow Taxi Trip Records dataset row-by-row and generates equivalent traces as follows. \\

\begin{table*}[t]
\centering
\caption{\label{Table1}Example of a trace file generated from a row in the dataset}
\scalebox{1}{
\begin{tabular}{|c|c|c|c|c|c|c|c|c|c|c|}
\hline
\begin{tabular}[c]{@{}c@{}}$time-$\\$stamp$ \end{tabular}& \begin{tabular}[c]{@{}c@{}}$req\_$\\$taxi$\end{tabular} & \begin{tabular}[c]{@{}c@{}}$wait\_$\\$time$\end{tabular} & \begin{tabular}[c]{@{}c@{}}$arrival\_$\\$loc$\end{tabular} & \begin{tabular}[c]{@{}c@{}}$pickup\_$\\$delay$\end{tabular} &\begin{tabular}[c]{@{}c@{}} $pick\_$\\$up$\end{tabular} & \begin{tabular}[c]{@{}c@{}}$drop\_$\\$loc$\end{tabular} & \begin{tabular}[c]{@{}c@{}}$fare\_$\\$amount$\end{tabular} & \begin{tabular}[c]{@{}c@{}}$trip\_$\\$distance$\end{tabular} &\begin{tabular}[c]{@{}c@{}}$rate\_$ \\$trip$\end{tabular} & \begin{tabular}[c]{@{}c@{}}$cong\_$\\ $charge$\end{tabular} \\ \hline
0  & $\top$ & - & $\bot$ & - & $\bot$ & $\bot$  & -  & - & -  & -  \\ \hline
1 &     & -    & $\bot$    & -  & $\bot$ & $\bot$ & - & - & - & -  \\ \hline
2 &     & 2 & $\top$   & -   & $\bot$  & $\bot$ & - & -  & -& -    \\ \hline
3 &   &   &   & -   & $\bot$ & $\bot$    & -  & -   & -    & -  \\ \hline
4  &  &   &   & 2   & $\top$     & $\bot$ & - & - & -  & -   \\ \hline
5  &  &  &   &  &  & $\bot$ & - & - & - & -  \\ \hline
6  &  &  &  &  & & $\bot$  & - & -   & -   & -  \\ \hline
7  &  &  &  & &  & $\bot$ & -  & - & -   & -   \\ \hline
8  &  &  &  &  &  & $\bot$ & -   & -  & -    & -  \\ \hline
9  &  &  &  &  &  & $\bot$ & -  & -   & -   & -  \\ \hline
10 &  &  &   &  & & $\top$ & 8  & 2.7 & 3  & 2.5 \\ \hline
\end{tabular}}
\end{table*}

\indent \textit{Trace generation.} To elucidate the trace generation process, let us consider the first row of the dataset (with augmented fields) as follows: $taxi\_request\_time$: 1/1/2021 12:26:00 AM, $arrival\_time$: 1/1/2021 12:28:00 AM, $wait\_time$: 2, $pickup\_delay$: 2, $tpep\_pickup\_$\\$datetime$: 1/1/2021 12:30:00 AM, $tpep\_dropoff\_datetime$: 1/1/2021  12:36:00 AM, $fare\_amount$:8, $trip\_distance$: 2.1, $ratings$: 3, and, $congestion\_charge$: 2.5. We ignore the other fields in the dataset while generating traces as we do not require them to monitor the properties presented in this paper. The generated trace file using the Python program is shown in Table~\ref{Table1} where $request\_taxi$, $arrival\_loc$, $pick\_up$ and $drop\_loc$ are atomic propositions denoting request for taxi service, arrival at the pickup location, picking up the passenger and dropping off passengers at the drop-off location.\\

\indent As shown in Table~\ref{Table1}, the time-stamps associated with the atomic propositions and parameters of interest are generated and uniquely assigned in a monotonically increasing order as the sequence of events progresses. The atomic proposition $req\_taxi$ is $\top$ at time-stamp 0, since that is the first proposition to be satisfied. The atomic proposition $arrival\_loc$ is $\top$ at time-stamp 2 because this proposition is satisfied 2 time units after the proposition  $req\_taxi$ is satisfied (i.e. $wait\_time$ = 2). Similarly, the atomic proposition $pick\_up$ is $\top$ at time-stamp 4, because the passenger was picked up 2 time units after the atomic proposition $arrival\_loc$ is satisfied (i.e. $pickup\_delay$ = 2). Similarly, the proposition $drop\_loc$ is $\top$ at time-stamp 10 because the proposition $drop\_loc$ was satisfied 6 time units after the proposition $pick\_up$ is $\top$. The other values in the trace file are populated in a similar way.\\

\indent \textit{Implementation.} \indent The Python program takes approximately $121.72$ seconds and consumed approximately $3.54$ MB memory for generating 100,000 traces from the dataset. The proposed monitoring algorithms are implemented using OCaml~\cite{minsky2013real} that accepts a finite set of traces and checks them against a given QTWTL specification. The QTWTL monitor returns $\top$ when a given set of traces $\TT$ satisfies the QTWTL specification $\phi$, and $\bot$ when the specification is violated. We monitor five QTWTL specifications denoted as $\phi_{1}--\phi_{5}$ as follows. \\

 \noindent \textbf{Requirement 1}: The maximum trip distance ($distance\_{trip}$) for a taxi driver should never cross 100 miles. \\
 \noindent \textit{QTWTL Specification 1}: $\phi_1 = \AAA_{max}({distance\_{trip}}) < 100$ \\



\noindent \textbf{Requirement 2}: If the average waiting time (${wait\_time}$) is less than 3 minutes then the average trip rating ($rate\_{trip}$) should be more than 3 stars. \\
\textit{QTWTL Specification 2}: $\phi_2 = \AAA_{avg}({wait\_time}) < 3 \longrightarrow\AAA_{avg}(rate\_$\\${trip}) > 3.$ \\





\noindent \textbf{Requirement 3}: At least $85\%$ of taxis should arrive at the pickup location ($arrival\_loc$) within 10 time units of requesting it ($request\_taxi$), pick-up ($pick\_up$) the customer within next 5 time units and proceed to the drop off ($drop\_loc$) location within next 35 time units. \\
\textit{QTWTL Specification 3}: $\phi_3 = \CC([\HH^{1}~request\_taxi] \longrightarrow  [\HH^{1}~arrival\_$\\$loc]^{[0, 10]} ~ \ser [\HH^{1}~pick\_up]^{[11, 15]} ~ \ser [\HH^{1}~drop\_loc]^{[16, 50]}) \geq 0.85$  \\

\noindent \textbf{Requirement 4}: If the average congestion charge ($congestion\_{charge}$) is less than \$2 and the average trip distance ($trip\_{distance}$) is less than 5 miles then the average fare amount ($fare\_{amount}$) should be less than \$10. \\
\textit{QTWTL Specification 4}: $\phi_4 = \AAA_{avg}(congestion\_{charge}) < 2 ~\land~$\\$  \AAA_{avg}(trip\_{distance})$ $< 5 \longrightarrow \AAA_{avg}(fare\_{amount}) < 10$  \\


\noindent \textbf{Requirement 5}: The minimum trip rating ($rate\_trip$) of taxis should be 3 stars if requirement $\phi_3$ is satisfied. \\
\textit{QTWTL Specification 5}: $\phi_5 = \phi_3 \longrightarrow \AAA_{min}({rate\_trip}) > 3.$ \\

\begin{figure*}[!t]
    \centering
    \subfloat[]{{\includegraphics[width=0.25\linewidth]{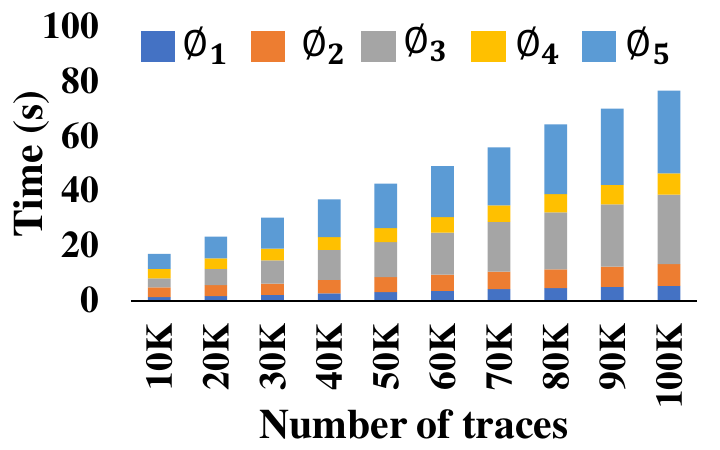}}}
    \subfloat[]{{\includegraphics[width=0.25\linewidth]{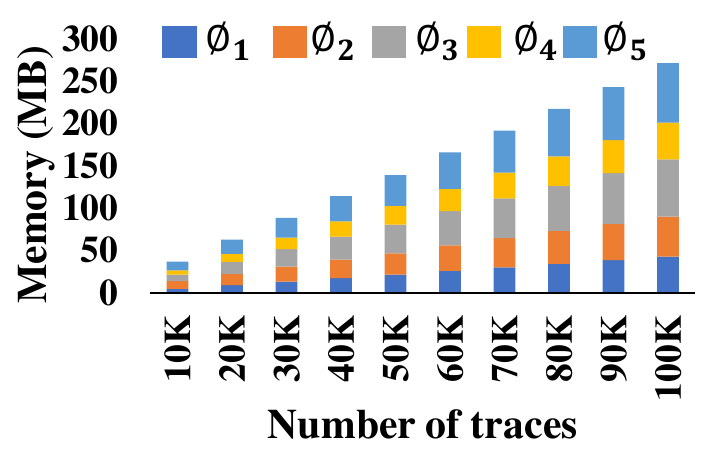}}} 
    \subfloat[]{{\includegraphics[width=0.25\linewidth]{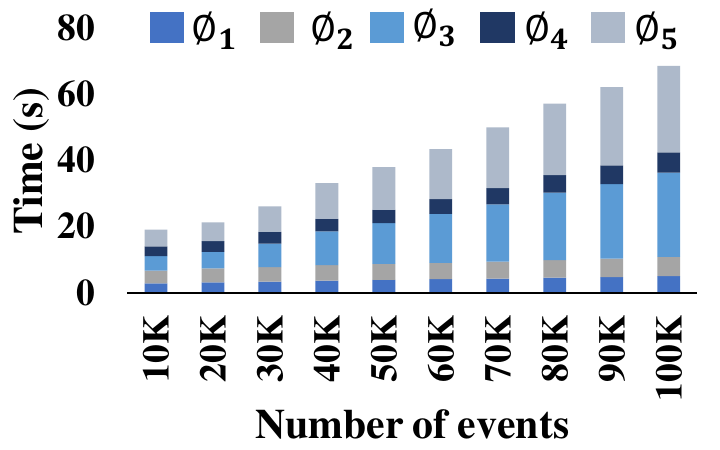}}} 
    \subfloat[]{{\includegraphics[width=0.25\linewidth]{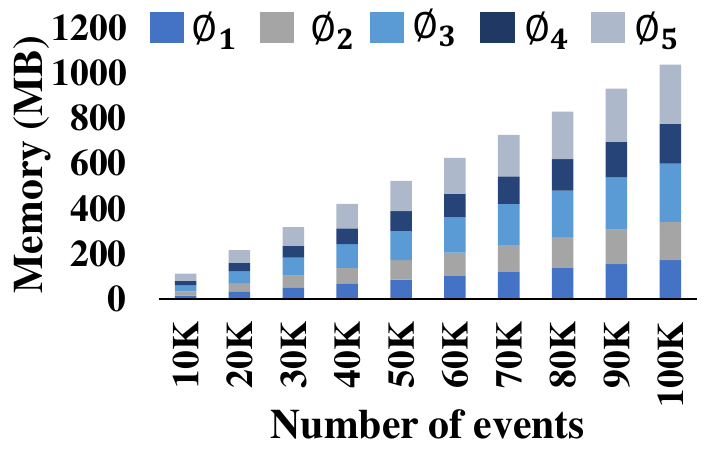}}} 
        \vspace{-2mm}
        \caption{(a) Number of traces vs. execution time  for $\phi_1$ - $\phi_5$; (b) Number of traces vs. memory consumption for $\phi_1$ - $\phi_5$; (c) Number of events vs. execution time for $\phi_1$ - $\phi_5$; (d) Number of events vs. memory consumption for $\phi_1$ - $\phi_5$}
        \label{fig:graphs1}
\end{figure*}

We performed two sets of experiments to analyze the relationship between the number of traces and performance and the relationship between the number of events and performance. In the first set of experiments, we evaluated the QTWTL formula, $\phi_1$ - $\phi_5$ against different numbers of traces ranging from 10,000 to 100,000 while keeping the number of events in each trace constant at 50. Their respective execution time and memory consumption were recorded. In the second set of experiments, we evaluated the formula $\phi_1$--$\phi_5$ against varying trace lengths ranging from 10,000 to 100,000 while keeping the number of traces constant at 200. The obtained results for the execution time and memory consumption for the first and second sets of experiments are shown in Figure ~\ref{fig:graphs1}. All the experiments were performed on Windows 10 Enterprise 64-bit system with 64 GB RAM and Intel Core(TM) i9-10900 CPU (3.70 GHz). \\ 

\indent \textcolor{black}{\textbf{Number of traces vs. performance:} We observe in Figure ~\ref{fig:graphs1}(a) that for each QTWTL formula, the execution time increases with an increase in the number of traces. For instance, for monitoring of $\phi_1$ for $10,000$ traces, our algorithm takes  $1.4$ seconds. The execution  time increases to $3.2$ seconds, $4.3$ seconds and $5.5$ during the monitoring of $\phi_1$ for $50,000$, $70,000$ and $100,000$ traces respectively. Similarly, our algorithm takes $5.5$ seconds for monitoring $\phi_5$ for $10,000$ traces. Once again, the execution time increases to $16.2$ seconds, $21.1$ seconds and $30.2$ seconds while monitoring $\phi_5$ for $50,000$, $70,000$ and $100,000$ traces respectively. It can be observed in Figure ~\ref{fig:graphs1}(b) that the memory consumption for QTWTL specifications also increases with the number of traces. For instance, while monitoring for $10,000$, $60,000$ and $100,000$ traces for $\phi_1$, the algorithm consumes $5.1$ MB, $26.7$ MB, and $43.1$ MB respectively. Once again, while monitoring $\phi_5$ for $10,000$, $60,000$ and $100,000$ traces, our algorithm consumes $10.2$ MB, $42.9$ MB, and $70.5$ MB, respectively. We observe a linear trend for both execution time and memory consumption.}\\

\indent \textcolor{black}{\textbf{Number of events vs. performance:} We observe in Figure ~\ref{fig:graphs1}(c) that the execution time increases with an increase in the number of events in the trace. For instance, for monitoring $\phi_1$ against 200 traces, each with $10,000$ events, the proposed algorithm takes $2.9$ seconds. The time however increases to $4.0$, $4.4$, and $5.2$ seconds for the same formula and a set of traces with $50,000$ $70,000$, and $100,000$ events respectively. Similarly, while monitoring $\phi_5$ for $10,000$ traces, our algorithm consumes $5.1$ seconds. Once again, the execution time increases to $12.9$, $18.5$, and $26.2$ seconds for the same formula and a set of traces with $50,000$, $70,000$ and $100,000$ events respectively. Also, we observe in Figure ~\ref{fig:graphs1}(d) that memory consumption increases with an increase in the number of events in the set of traces. The memory consumed for monitoring $\phi_1$ for $200$ traces each with $10,000$ events is $17.2$ MB. However, the memory consumed increases to $87.9$ MB and $174.0$ MB while monitoring $\phi_1$ for $200$ traces each with $50,000$ and $100,000$ events respectively. Similarly, the memory consumed for monitoring $\phi_4$ for $200$ traces each with $10,000$ events is $19.2$ MB. However, the memory consumed increases to $87.9$ MB and $177.0$ MB while monitoring $\phi_4$ for $200$ traces each with $50,000$ and $100,000$ events, respectively.} \\

\indent The obtained results from the case study show the effectiveness and efficiency of our proposed QTWTL runtime monitoring algorithm. We observe that monitoring algorithm has a linear space and time complexity with respect to the number of traces monitored. Even though the illustration used logged traces, the same monitor can be also used for online monitoring with some modifications in the algorithm. We believe that such monitoring can be used for service-based agents by inspecting the logged traces and by adding monitors in the early controller prototypes for quickly evaluating not only their safety but also their performance requirements. Thus, engineers working on the design and development of the service-based agents can use the proposed approach with limited, or even no special knowledge of runtime verification and gain formally verified performance-related insights about the given robotic system. On the other hand, the proposed approach can be used by the optimization engineers as well to enhance the quality of service of deployed agents in different applications such as autonomous delivery, transportation cleaning, manufacturing, and many others. 

\section{Related Works}
\label{related_works}
In recent times, various conventional temporal formalisms have been extended with various aggregation modalities in different applications. In \cite{sistla1995temporal}, Past Temporal Logic (PTL) was extended with temporal aggregate functions for specifying events and conditions in active database systems over time. The authors then presented an incremental algorithm for evaluating the satisfaction or violation of formulae specified in the language. Linear-time temporal logic was extended in \cite{finkbeiner2002collecting} with aggregation operators for expressing temporal specifications over a collection of statistical data. The specifications are then translated into queries which are subsequently evaluated using an automata-based query evaluation algorithm proposed by the authors. In \cite{bianculli2014trace}, Metric Temporal Logic was extended with aggregating modalities for formalizing service level agreement requirements in service-based applications. The authors further proposed a model checking approach based on MapReduce to evaluate the specifications expressed in this extended logic. Inspired by \cite{hella2001logics}, the authors in \cite{mtfol} extended Metric First-Order Temporal Logic (MFOTL) with SQL-like aggregation operators. The authors also proposed a monitoring algorithm to evaluate database queries captured as MFOTL specifications. In \cite{basin2017monpoly}, MonPoly, another monitoring tool for checking traces against formulae specified using MFOTL, was also proposed. 
Despite the expressiveness of the extended MFOTL in formalizing compliance policies, the absence of an explicit concatenation operator in its semantics limits its expression of serial tasks in real-world applications. However, this limitation can be addressed using the compact semantics of QTWTL. Furthermore, following various extensions to Signal Temporal Logic with aggregation operators to define spatial-temporal requirements in CPS applications in \cite{haghighi2015spatel, ma2018cityresolver, maler2004monitoring, nenzi2015qualitative, massink2018qualitative, bartocci2017monitoring}, Spatial Aggregation Signal Signal Temporal Logic (SaSTL) which extends STL with two operators for expressing spatial aggregation and spatial counting characteristics in smart city requirements was proposed in \cite{ma2020sastl}. In comparison to these works, as previously mentioned in the introduction, classical TWTL formalism has several advantages, including compactness obtained through the use of the concatenation operator ($\ser$), which is very useful in robotic applications. Our proposed QTWTL formalism leverages the compactness of classical TWTL and extends the logic with two new operators. It is worth mentioning that recently TWTL was extended with quantifiers and trace variables to specify synchronous hyperproperties in \cite{10137643} and using an SMT solver, a security-aware robotic motion planning approach was proposed. Furthermore, model checking algorithm for both synchronous and asynchronous HyperTWTL was proposed in \cite{bonnah2023model}.


\section{Conclusion}
\label{conclusion}
In this paper, we proposed a novel specification language quality aware time window temporal logic (QTWTL) by extending the time window temporal logic (TWTL) with counting and aggregation operators. We also presented offline runtime monitoring algorithms for monitoring the performance of service-based agents, which is indeed an important metric for assuring their quality of service. To demonstrate the effectiveness of our proposed approach, we generated a large number of traces using the real-world New York City Taxi and Limousine Commission Trip Record dataset, formalized performance requirements using QTWTL, and monitored them using the proposed algorithms.  
The obtained results show that the proposed monitoring approach is efficient in terms of memory and execution time with an increasing number of traces. In the future, we plan to extend the QTWTL specification language with new temporal operators and provide online runtime monitoring algorithms for them. Also, we plan to implement the online version of the proposed algorithms on robots and demonstrate their on-field effectiveness.

\bibliographystyle{ACM-Reference-Format}
\bibliography{Ref}


\begin{thebibliography}{42}


\ifx \showCODEN    \undefined \def \showCODEN     #1{\unskip}     \fi
\ifx \showDOI      \undefined \def \showDOI       #1{#1}\fi
\ifx \showISBNx    \undefined \def \showISBNx     #1{\unskip}     \fi
\ifx \showISBNxiii \undefined \def \showISBNxiii  #1{\unskip}     \fi
\ifx \showISSN     \undefined \def \showISSN      #1{\unskip}     \fi
\ifx \showLCCN     \undefined \def \showLCCN      #1{\unskip}     \fi
\ifx \shownote     \undefined \def \shownote      #1{#1}          \fi
\ifx \showarticletitle \undefined \def \showarticletitle #1{#1}   \fi
\ifx \showURL      \undefined \def \showURL       {\relax}        \fi
\providecommand\bibfield[2]{#2}
\providecommand\bibinfo[2]{#2}
\providecommand\natexlab[1]{#1}
\providecommand\showeprint[2][]{arXiv:#2}

\bibitem[Aksaray et~al\mbox{.}(2016)]%
        {re3}
\bibfield{author}{\bibinfo{person}{Derya Aksaray},
  \bibinfo{person}{Cristian-Ioan Vasile}, {and} \bibinfo{person}{Calin Belta}.}
  \bibinfo{year}{2016}\natexlab{}.
\newblock \showarticletitle{Dynamic routing of energy-aware vehicles with
  Temporal Logic Constraints}. In \bibinfo{booktitle}{\emph{2016 IEEE
  International Conference on Robotics and Automation (ICRA)}}.
  \bibinfo{pages}{3141--3146}.
\newblock
\urldef\tempurl%
\url{https://doi.org/10.1109/ICRA.2016.7487481}
\showDOI{\tempurl}


\bibitem[Asarkaya et~al\mbox{.}(2021)]%
        {re1}
\bibfield{author}{\bibinfo{person}{Ahmet~Semi Asarkaya}, \bibinfo{person}{Derya
  Aksaray}, {and} \bibinfo{person}{Yasin Yaz{\i}c{\i}o{\u{g}}lu}.}
  \bibinfo{year}{2021}\natexlab{}.
\newblock \showarticletitle{Temporal-Logic-Constrained Hybrid Reinforcement
  Learning to Perform Optimal Aerial Monitoring with Delivery Drones}. In
  \bibinfo{booktitle}{\emph{2021 International Conference on Unmanned Aircraft
  Systems (ICUAS)}}. IEEE, \bibinfo{pages}{285--294}.
\newblock


\bibitem[Attard et~al\mbox{.}(2021)]%
        {attard2021better}
\bibfield{author}{\bibinfo{person}{Duncan~Paul Attard}, \bibinfo{person}{Luca
  Aceto}, \bibinfo{person}{Antonis Achilleos}, \bibinfo{person}{Adrian
  Francalanza}, \bibinfo{person}{Anna Ing{\'o}lfsd{\'o}ttir}, {and}
  \bibinfo{person}{Karoliina Lehtinen}.} \bibinfo{year}{2021}\natexlab{}.
\newblock \showarticletitle{Better late than never or: verifying asynchronous
  components at runtime}. In \bibinfo{booktitle}{\emph{Formal Techniques for
  Distributed Objects, Components, and Systems: 41st IFIP WG 6.1 International
  Conference, FORTE 2021, Held as Part of the 16th International Federated
  Conference on Distributed Computing Techniques, DisCoTec 2021, Valletta,
  Malta, June 14--18, 2021, Proceedings}}. Springer, \bibinfo{pages}{207--225}.
\newblock


\bibitem[Bartocci et~al\mbox{.}(2017)]%
        {bartocci2017monitoring}
\bibfield{author}{\bibinfo{person}{Ezio Bartocci}, \bibinfo{person}{Luca
  Bortolussi}, \bibinfo{person}{Michele Loreti}, {and} \bibinfo{person}{Laura
  Nenzi}.} \bibinfo{year}{2017}\natexlab{}.
\newblock \showarticletitle{Monitoring mobile and spatially distributed
  cyber-physical systems}. In \bibinfo{booktitle}{\emph{Proceedings of the 15th
  ACM-IEEE International Conference on Formal Methods and Models for System
  Design}}. \bibinfo{pages}{146--155}.
\newblock


\bibitem[Basin et~al\mbox{.}(2015)]%
        {mtfol}
\bibfield{author}{\bibinfo{person}{David Basin}, \bibinfo{person}{Felix
  Klaedtke}, \bibinfo{person}{Srdjan Marinovic}, {and} \bibinfo{person}{Eugen
  Z\u{a}linescu}.} \bibinfo{year}{2015}\natexlab{}.
\newblock \showarticletitle{Monitoring of Temporal First-Order Properties with
  Aggregations}.
\newblock \bibinfo{journal}{\emph{Form. Methods Syst. Des.}}
  \bibinfo{volume}{46}, \bibinfo{number}{3} (\bibinfo{date}{June}
  \bibinfo{year}{2015}), \bibinfo{pages}{262–285}.
\newblock
\showISSN{0925-9856}
\urldef\tempurl%
\url{https://doi.org/10.1007/s10703-015-0222-7}
\showDOI{\tempurl}


\bibitem[Basin et~al\mbox{.}(2017)]%
        {basin2017monpoly}
\bibfield{author}{\bibinfo{person}{David~A Basin}, \bibinfo{person}{Felix
  Klaedtke}, {and} \bibinfo{person}{Eugen Zalinescu}.}
  \bibinfo{year}{2017}\natexlab{}.
\newblock \showarticletitle{The MonPoly Monitoring Tool.}
\newblock \bibinfo{journal}{\emph{RV-CuBES}}  \bibinfo{volume}{3}
  (\bibinfo{year}{2017}), \bibinfo{pages}{19--28}.
\newblock


\bibitem[Bianculli et~al\mbox{.}(2014)]%
        {bianculli2014trace}
\bibfield{author}{\bibinfo{person}{Domenico Bianculli}, \bibinfo{person}{Carlo
  Ghezzi}, {and} \bibinfo{person}{Sr{\dj}an Krsti{\'c}}.}
  \bibinfo{year}{2014}\natexlab{}.
\newblock \showarticletitle{Trace checking of metric temporal logic with
  aggregating modalities using MapReduce}. In
  \bibinfo{booktitle}{\emph{International Conference on Software Engineering
  and Formal Methods}}. Springer, \bibinfo{pages}{144--158}.
\newblock


\bibitem[Bonnah and Hoque(2022)]%
        {bonnah2022runtime}
\bibfield{author}{\bibinfo{person}{Ernest Bonnah} {and}
  \bibinfo{person}{Khaza~Anuarul Hoque}.} \bibinfo{year}{2022}\natexlab{}.
\newblock \showarticletitle{Runtime Monitoring of Time Window Temporal Logic}.
\newblock \bibinfo{journal}{\emph{IEEE Robotics and Automation Letters}}
  \bibinfo{volume}{7}, \bibinfo{number}{3} (\bibinfo{year}{2022}),
  \bibinfo{pages}{5888--5895}.
\newblock


\bibitem[Bonnah et~al\mbox{.}(2023a)]%
        {10137643}
\bibfield{author}{\bibinfo{person}{Ernest Bonnah}, \bibinfo{person}{Luan
  Nguyen}, {and} \bibinfo{person}{Khaza~Anuarul Hoque}.}
  \bibinfo{year}{2023}\natexlab{a}.
\newblock \showarticletitle{Motion Planning Using Hyperproperties for Time
  Window Temporal Logic}.
\newblock \bibinfo{journal}{\emph{IEEE Robotics and Automation Letters}}
  \bibinfo{volume}{8}, \bibinfo{number}{8} (\bibinfo{year}{2023}),
  \bibinfo{pages}{4386--4393}.
\newblock
\urldef\tempurl%
\url{https://doi.org/10.1109/LRA.2023.3280830}
\showDOI{\tempurl}


\bibitem[Bonnah et~al\mbox{.}(2023b)]%
        {bonnah2023model}
\bibfield{author}{\bibinfo{person}{Ernest Bonnah}, \bibinfo{person}{Luan~Viet
  Nguyen}, {and} \bibinfo{person}{Khaza~Anuarul Hoque}.}
  \bibinfo{year}{2023}\natexlab{b}.
\newblock \bibinfo{title}{Model Checking Time Window Temporal Logic for
  Hyperproperties}.
\newblock
\newblock
\showeprint[arxiv]{2308.02554}~[cs.LO]


\bibitem[Cai et~al\mbox{.}(2021)]%
        {cai2021modular}
\bibfield{author}{\bibinfo{person}{Mingyu Cai}, \bibinfo{person}{Mohammadhosein
  Hasanbeig}, \bibinfo{person}{Shaoping Xiao}, \bibinfo{person}{Alessandro
  Abate}, {and} \bibinfo{person}{Zhen Kan}.} \bibinfo{year}{2021}\natexlab{}.
\newblock \showarticletitle{Modular deep reinforcement learning for continuous
  motion planning with temporal logic}.
\newblock \bibinfo{journal}{\emph{IEEE Robotics and Automation Letters}}
  \bibinfo{volume}{6}, \bibinfo{number}{4} (\bibinfo{year}{2021}),
  \bibinfo{pages}{7973--7980}.
\newblock


\bibitem[Cassar et~al\mbox{.}(2017)]%
        {off4}
\bibfield{author}{\bibinfo{person}{Ian Cassar}, \bibinfo{person}{Adrian
  Francalanza}, \bibinfo{person}{Luca Aceto}, {and} \bibinfo{person}{Anna
  Ing{\'o}lfsd{\'o}ttir}.} \bibinfo{year}{2017}\natexlab{}.
\newblock \showarticletitle{A survey of runtime monitoring instrumentation
  techniques}.
\newblock \bibinfo{journal}{\emph{arXiv preprint arXiv:1708.07229}}
  (\bibinfo{year}{2017}).
\newblock


\bibitem[Chen and Ro{\c{s}}u(2007)]%
        {off1}
\bibfield{author}{\bibinfo{person}{Feng Chen} {and} \bibinfo{person}{Grigore
  Ro{\c{s}}u}.} \bibinfo{year}{2007}\natexlab{}.
\newblock \showarticletitle{Mop: an efficient and generic runtime verification
  framework}. In \bibinfo{booktitle}{\emph{Proceedings of the 22nd annual ACM
  SIGPLAN conference on Object-oriented programming systems, languages and
  applications}}. \bibinfo{pages}{569--588}.
\newblock


\bibitem[Finkbeiner et~al\mbox{.}(2002)]%
        {finkbeiner2002collecting}
\bibfield{author}{\bibinfo{person}{Bernd Finkbeiner}, \bibinfo{person}{Sriram
  Sankaranarayanan}, {and} \bibinfo{person}{Henny Sipma}.}
  \bibinfo{year}{2002}\natexlab{}.
\newblock \showarticletitle{Collecting statistics over runtime executions}.
\newblock \bibinfo{journal}{\emph{Electronic Notes in Theoretical Computer
  Science}} \bibinfo{volume}{70}, \bibinfo{number}{4} (\bibinfo{year}{2002}),
  \bibinfo{pages}{36--54}.
\newblock


\bibitem[Guglielmetti~Mugion et~al\mbox{.}(2022)]%
        {guglielmetti2022improving}
\bibfield{author}{\bibinfo{person}{Roberta Guglielmetti~Mugion},
  \bibinfo{person}{Maria~Francesca Renzi}, {and} \bibinfo{person}{Laura
  Di~Pietro}.} \bibinfo{year}{2022}\natexlab{}.
\newblock \showarticletitle{Improving Service Quality Through Individuals’
  Satisfaction. Evidence from the Healthcare Sector}.
\newblock In \bibinfo{booktitle}{\emph{The Palgrave Handbook of Service
  Management}}. \bibinfo{publisher}{Springer}, \bibinfo{pages}{745--772}.
\newblock


\bibitem[Haghighi et~al\mbox{.}(2015)]%
        {haghighi2015spatel}
\bibfield{author}{\bibinfo{person}{Iman Haghighi}, \bibinfo{person}{Austin
  Jones}, \bibinfo{person}{Zhaodan Kong}, \bibinfo{person}{Ezio Bartocci},
  \bibinfo{person}{Radu Gros}, {and} \bibinfo{person}{Calin Belta}.}
  \bibinfo{year}{2015}\natexlab{}.
\newblock \showarticletitle{SpaTeL: a novel spatial-temporal logic and its
  applications to networked systems}. In \bibinfo{booktitle}{\emph{Proceedings
  of the 18th International Conference on Hybrid Systems: Computation and
  Control}}. \bibinfo{pages}{189--198}.
\newblock


\bibitem[Hella et~al\mbox{.}(2001)]%
        {hella2001logics}
\bibfield{author}{\bibinfo{person}{Lauri Hella}, \bibinfo{person}{Leonid
  Libkin}, \bibinfo{person}{Juha Nurmonen}, {and} \bibinfo{person}{Limsoon
  Wong}.} \bibinfo{year}{2001}\natexlab{}.
\newblock \showarticletitle{Logics with aggregate operators}.
\newblock \bibinfo{journal}{\emph{Journal of the ACM (JACM)}}
  \bibinfo{volume}{48}, \bibinfo{number}{4} (\bibinfo{year}{2001}),
  \bibinfo{pages}{880--907}.
\newblock


\bibitem[Kloetzer and Mahulea(2020)]%
        {kloetzer2020path}
\bibfield{author}{\bibinfo{person}{Marius Kloetzer} {and}
  \bibinfo{person}{Cristian Mahulea}.} \bibinfo{year}{2020}\natexlab{}.
\newblock \showarticletitle{Path planning for robotic teams based on LTL
  specifications and Petri net models}.
\newblock \bibinfo{journal}{\emph{Discrete Event Dynamic Systems}}
  \bibinfo{volume}{30}, \bibinfo{number}{1} (\bibinfo{year}{2020}),
  \bibinfo{pages}{55--79}.
\newblock


\bibitem[Leucker and Schallhart(2009)]%
        {leucker2009brief}
\bibfield{author}{\bibinfo{person}{Martin Leucker} {and}
  \bibinfo{person}{Christian Schallhart}.} \bibinfo{year}{2009}\natexlab{}.
\newblock \showarticletitle{A brief account of runtime verification}.
\newblock \bibinfo{journal}{\emph{The Journal of Logic and Algebraic
  Programming}} \bibinfo{volume}{78}, \bibinfo{number}{5}
  (\bibinfo{year}{2009}), \bibinfo{pages}{293--303}.
\newblock


\bibitem[Ma et~al\mbox{.}(2020)]%
        {ma2020sastl}
\bibfield{author}{\bibinfo{person}{Meiyi Ma}, \bibinfo{person}{Ezio Bartocci},
  \bibinfo{person}{Eli Lifland}, \bibinfo{person}{John Stankovic}, {and}
  \bibinfo{person}{Lu Feng}.} \bibinfo{year}{2020}\natexlab{}.
\newblock \showarticletitle{SaSTL: Spatial aggregation signal temporal logic
  for runtime monitoring in smart cities}. In \bibinfo{booktitle}{\emph{2020
  ACM/IEEE 11th International Conference on Cyber-Physical Systems (ICCPS)}}.
  IEEE, \bibinfo{pages}{51--62}.
\newblock


\bibitem[Ma et~al\mbox{.}(2017)]%
        {ma2017runtime}
\bibfield{author}{\bibinfo{person}{Meiyi Ma}, \bibinfo{person}{John~A
  Stankovic}, {and} \bibinfo{person}{Lu Feng}.}
  \bibinfo{year}{2017}\natexlab{}.
\newblock \showarticletitle{Runtime monitoring of safety and performance
  requirements in smart cities}. In \bibinfo{booktitle}{\emph{Proceedings of
  the 1st ACM Workshop on the Internet of Safe Things}}.
  \bibinfo{pages}{44--50}.
\newblock


\bibitem[Ma et~al\mbox{.}(2018)]%
        {ma2018cityresolver}
\bibfield{author}{\bibinfo{person}{Meiyi Ma}, \bibinfo{person}{John~A
  Stankovic}, {and} \bibinfo{person}{Lu Feng}.}
  \bibinfo{year}{2018}\natexlab{}.
\newblock \showarticletitle{Cityresolver: a decision support system for
  conflict resolution in smart cities}. In \bibinfo{booktitle}{\emph{2018
  ACM/IEEE 9th International Conference on Cyber-Physical Systems (ICCPS)}}.
  IEEE, \bibinfo{pages}{55--64}.
\newblock


\bibitem[Maler and Nickovic(2004)]%
        {maler2004monitoring}
\bibfield{author}{\bibinfo{person}{Oded Maler} {and} \bibinfo{person}{Dejan
  Nickovic}.} \bibinfo{year}{2004}\natexlab{}.
\newblock \showarticletitle{Monitoring temporal properties of continuous
  signals}.
\newblock In \bibinfo{booktitle}{\emph{Formal Techniques, Modelling and
  Analysis of Timed and Fault-Tolerant Systems}}.
  \bibinfo{publisher}{Springer}, \bibinfo{pages}{152--166}.
\newblock


\bibitem[Massink et~al\mbox{.}(2018)]%
        {massink2018qualitative}
\bibfield{author}{\bibinfo{person}{M Massink}, \bibinfo{person}{M Loreti},
  \bibinfo{person}{Vincenzo Ciancia}, \bibinfo{person}{L Bortolussi}, {and}
  \bibinfo{person}{L Nenzi}.} \bibinfo{year}{2018}\natexlab{}.
\newblock \showarticletitle{Qualitative and quantitative monitoring of
  spatio-temporal properties with SSTL}.
\newblock \bibinfo{journal}{\emph{Logical Methods in Computer Science}}
  \bibinfo{volume}{14} (\bibinfo{year}{2018}).
\newblock


\bibitem[Mendes and Guerreiro(2016)]%
        {mendes2016monitoring}
\bibfield{author}{\bibinfo{person}{J{\'u}lio Mendes} {and}
  \bibinfo{person}{Manuela Guerreiro}.} \bibinfo{year}{2016}\natexlab{}.
\newblock \showarticletitle{Monitoring the quality of tourism experience}. In
  \bibinfo{booktitle}{\emph{Asia Tourism Forum 2016-the 12th Biennial
  Conference of Hospitality and Tourism Industry in Asia}}. Atlantis Press,
  \bibinfo{pages}{300--310}.
\newblock


\bibitem[Meredith and Ro{\c{s}}u(2010)]%
        {off3}
\bibfield{author}{\bibinfo{person}{Patrick Meredith} {and}
  \bibinfo{person}{Grigore Ro{\c{s}}u}.} \bibinfo{year}{2010}\natexlab{}.
\newblock \showarticletitle{Runtime verification with the RV system}. In
  \bibinfo{booktitle}{\emph{International Conference on Runtime Verification}}.
  Springer, \bibinfo{pages}{136--152}.
\newblock


\bibitem[Minsky et~al\mbox{.}(2013)]%
        {minsky2013real}
\bibfield{author}{\bibinfo{person}{Yaron Minsky}, \bibinfo{person}{Anil
  Madhavapeddy}, {and} \bibinfo{person}{Jason Hickey}.}
  \bibinfo{year}{2013}\natexlab{}.
\newblock \bibinfo{booktitle}{\emph{Real World OCaml: Functional programming
  for the masses}}.
\newblock \bibinfo{publisher}{" O'Reilly Media, Inc."}.
\newblock


\bibitem[Mosca et~al\mbox{.}(2019)]%
        {mosca2019multi}
\bibfield{author}{\bibinfo{person}{Alessio Mosca},
  \bibinfo{person}{Cristian-Ioan Vasile}, \bibinfo{person}{Calin Belta}, {and}
  \bibinfo{person}{Davide~M Raimondo}.} \bibinfo{year}{2019}\natexlab{}.
\newblock \showarticletitle{Multi-robot routing and scheduling with temporal
  logic and synchronization constraints}. In
  \bibinfo{booktitle}{\emph{Proceedings of the 2019 2nd International
  Conference on Control and Robot Technology}}. \bibinfo{pages}{40--45}.
\newblock


\bibitem[Nenzi et~al\mbox{.}(2015)]%
        {nenzi2015qualitative}
\bibfield{author}{\bibinfo{person}{Laura Nenzi}, \bibinfo{person}{Luca
  Bortolussi}, \bibinfo{person}{Vincenzo Ciancia}, \bibinfo{person}{Michele
  Loreti}, {and} \bibinfo{person}{Mieke Massink}.}
  \bibinfo{year}{2015}\natexlab{}.
\newblock \showarticletitle{Qualitative and quantitative monitoring of
  spatio-temporal properties}. In \bibinfo{booktitle}{\emph{Runtime
  Verification}}. Springer, \bibinfo{pages}{21--37}.
\newblock


\bibitem[{NYC Taxi and Limousine Commission (TLC)}(2022)]%
        {dataset}
\bibfield{author}{\bibinfo{person}{{NYC Taxi and Limousine Commission (TLC)}}.}
  \bibinfo{year}{2022}\natexlab{}.
\newblock \bibinfo{title}{{TLC Trip Record Data}}.
\newblock
  \bibinfo{howpublished}{\url{https://www1.nyc.gov/site/tlc/about/tlc-trip-record-data.page}}.
\newblock
\newblock
\shownote{[Online; accessed 29-Feb-2022]}.


\bibitem[Peterson et~al\mbox{.}(2020)]%
        {re2}
\bibfield{author}{\bibinfo{person}{Ryan Peterson}, \bibinfo{person}{Ali~Tevfik
  Buyukkocak}, \bibinfo{person}{Derya Aksaray}, {and} \bibinfo{person}{Yasin
  Yaz{\i}c{\i}oglu}.} \bibinfo{year}{2020}\natexlab{}.
\newblock \showarticletitle{Decentralized safe reactive planning under TWTL
  specifications}. In \bibinfo{booktitle}{\emph{2020 IEEE/RSJ International
  Conference on Intelligent Robots and Systems (IROS)}}. IEEE,
  \bibinfo{pages}{6599--6604}.
\newblock


\bibitem[Ro{\c{s}}u and Havelund(2005)]%
        {rocsu2005rewriting}
\bibfield{author}{\bibinfo{person}{Grigore Ro{\c{s}}u} {and}
  \bibinfo{person}{Klaus Havelund}.} \bibinfo{year}{2005}\natexlab{}.
\newblock \showarticletitle{Rewriting-based techniques for runtime
  verification}.
\newblock \bibinfo{journal}{\emph{Automated Software Engineering}}
  \bibinfo{volume}{12}, \bibinfo{number}{2} (\bibinfo{year}{2005}),
  \bibinfo{pages}{151--197}.
\newblock


\bibitem[Saha and Julius(2017)]%
        {saha2017task}
\bibfield{author}{\bibinfo{person}{Sayan Saha} {and}
  \bibinfo{person}{Anak~Agung Julius}.} \bibinfo{year}{2017}\natexlab{}.
\newblock \showarticletitle{Task and motion planning for manipulator arms with
  metric temporal logic specifications}.
\newblock \bibinfo{journal}{\emph{IEEE robotics and automation letters}}
  \bibinfo{volume}{3}, \bibinfo{number}{1} (\bibinfo{year}{2017}),
  \bibinfo{pages}{379--386}.
\newblock


\bibitem[Sahin et~al\mbox{.}(2019)]%
        {sahin2019multirobot}
\bibfield{author}{\bibinfo{person}{Yunus~Emre Sahin}, \bibinfo{person}{Petter
  Nilsson}, {and} \bibinfo{person}{Necmiye Ozay}.}
  \bibinfo{year}{2019}\natexlab{}.
\newblock \showarticletitle{Multirobot coordination with counting temporal
  logics}.
\newblock \bibinfo{journal}{\emph{IEEE Transactions on Robotics}}
  \bibinfo{volume}{36}, \bibinfo{number}{4} (\bibinfo{year}{2019}),
  \bibinfo{pages}{1189--1206}.
\newblock


\bibitem[S{\'a}nchez et~al\mbox{.}(2019)]%
        {sanchez2019survey}
\bibfield{author}{\bibinfo{person}{C{\'e}sar S{\'a}nchez},
  \bibinfo{person}{Gerardo Schneider}, \bibinfo{person}{Wolfgang Ahrendt},
  \bibinfo{person}{Ezio Bartocci}, \bibinfo{person}{Domenico Bianculli},
  \bibinfo{person}{Christian Colombo}, \bibinfo{person}{Yli{\`e}s Falcone},
  \bibinfo{person}{Adrian Francalanza}, \bibinfo{person}{Sr{\dj}an Krsti{\'c}},
  \bibinfo{person}{Joao~M Louren{\c{c}}o}, {et~al\mbox{.}}}
  \bibinfo{year}{2019}\natexlab{}.
\newblock \showarticletitle{A survey of challenges for runtime verification
  from advanced application domains (beyond software)}.
\newblock \bibinfo{journal}{\emph{Formal Methods in System Design}}
  \bibinfo{volume}{54}, \bibinfo{number}{3} (\bibinfo{year}{2019}),
  \bibinfo{pages}{279--335}.
\newblock


\bibitem[Shahzaad and Bouguettaya(2022)]%
        {shahzaad2022service}
\bibfield{author}{\bibinfo{person}{Babar Shahzaad} {and}
  \bibinfo{person}{Athman Bouguettaya}.} \bibinfo{year}{2022}\natexlab{}.
\newblock \showarticletitle{Service-Oriented Architecture for Drone-based
  Multi-Package Delivery}.
\newblock \bibinfo{journal}{\emph{arXiv preprint arXiv:2206.04544}}
  (\bibinfo{year}{2022}).
\newblock


\bibitem[Sistla and Wolfson(1995)]%
        {sistla1995temporal}
\bibfield{author}{\bibinfo{person}{A~Prasad Sistla} {and} \bibinfo{person}{Ouri
  Wolfson}.} \bibinfo{year}{1995}\natexlab{}.
\newblock \showarticletitle{Temporal conditions and integrity constraints in
  active database systems}. In \bibinfo{booktitle}{\emph{Proceedings of the
  1995 ACM SIGMOD International Conference on Management of Data}}.
  \bibinfo{pages}{269--280}.
\newblock


\bibitem[Sodhro et~al\mbox{.}(2019)]%
        {sodhro2019quality}
\bibfield{author}{\bibinfo{person}{Ali~Hassan Sodhro},
  \bibinfo{person}{Mohammad~S Obaidat}, \bibinfo{person}{Qammer~H Abbasi},
  \bibinfo{person}{Pasquale Pace}, \bibinfo{person}{Sandeep Pirbhulal},
  \bibinfo{person}{Giancarlo Fortino}, \bibinfo{person}{Muhammad~Ali Imran},
  \bibinfo{person}{Marwa Qaraqe}, {et~al\mbox{.}}}
  \bibinfo{year}{2019}\natexlab{}.
\newblock \showarticletitle{Quality of service optimization in an IoT-driven
  intelligent transportation system}.
\newblock \bibinfo{journal}{\emph{IEEE Wireless Communications}}
  \bibinfo{volume}{26}, \bibinfo{number}{6} (\bibinfo{year}{2019}),
  \bibinfo{pages}{10--17}.
\newblock


\bibitem[Tkachev and Abate(2013)]%
        {tkachev2013formula}
\bibfield{author}{\bibinfo{person}{Ilya Tkachev} {and}
  \bibinfo{person}{Alessandro Abate}.} \bibinfo{year}{2013}\natexlab{}.
\newblock \showarticletitle{Formula-free finite abstractions for linear
  temporal verification of stochastic hybrid systems}. In
  \bibinfo{booktitle}{\emph{Proceedings of the 16th international conference on
  Hybrid systems: computation and control}}. \bibinfo{pages}{283--292}.
\newblock


\bibitem[Vasile et~al\mbox{.}(2017)]%
        {vasile2017time}
\bibfield{author}{\bibinfo{person}{Cristian-Ioan Vasile},
  \bibinfo{person}{Derya Aksaray}, {and} \bibinfo{person}{Calin Belta}.}
  \bibinfo{year}{2017}\natexlab{}.
\newblock \showarticletitle{Time window temporal logic}.
\newblock \bibinfo{journal}{\emph{Theoretical Computer Science}}
  \bibinfo{volume}{691} (\bibinfo{year}{2017}), \bibinfo{pages}{27--54}.
\newblock


\bibitem[Xie et~al\mbox{.}(2020)]%
        {xie2020temporal}
\bibfield{author}{\bibinfo{person}{Guoshan Xie}, \bibinfo{person}{Zhihong Yin},
  {and} \bibinfo{person}{Jianqing Li}.} \bibinfo{year}{2020}\natexlab{}.
\newblock \showarticletitle{Temporal logic based motion planning with
  infeasible {LTL} specification}. In \bibinfo{booktitle}{\emph{2020 Chinese
  Control And Decision Conference (CCDC)}}. IEEE, \bibinfo{pages}{4899--4904}.
\newblock


\bibitem[Zhao et~al\mbox{.}(2022)]%
        {zhao2022astl}
\bibfield{author}{\bibinfo{person}{Deng Zhao}, \bibinfo{person}{Zhangbing
  Zhou}, \bibinfo{person}{Zhipeng Cai}, \bibinfo{person}{Sami Yangui}, {and}
  \bibinfo{person}{Xiao Xue}.} \bibinfo{year}{2022}\natexlab{}.
\newblock \showarticletitle{ASTL: Accumulative STL with a Novel Robustness
  Metric for IoT Service Monitoring}.
\newblock \bibinfo{journal}{\emph{IEEE Transactions on Mobile Computing}}
  (\bibinfo{year}{2022}).
\newblock


\end{thebibliography}

\end{document}